\documentclass[useAMS,usenatbib,referee]{mn2e}
\usepackage{epsf,graphicx,subfigure,lscape,longtable,txfonts}
\usepackage{rotating}
\usepackage{amssymb}
\usepackage[light]{}
\def\kms{\ifmmode{\rm km\,s}^{-1}\, \else km\,s$^{-1}$\fi\,}
\def\mujybm{$\rm{\mu}$Jy\,beam$^{-1}$\,}
\def\solmas{$\rm{M_\odot}$}

\def\HII{H{\sc {ii}}}

\begin{document}

\setlength{\pdfpageheight}{\paperheight}
\setlength{\pdfpagewidth}{\paperwidth}
  
\title[MERLIN 5\,GHz Radio Imaging of SNR in M82]{Deep MERLIN 5\,GHz Radio Imaging of Supernova Remnants in the M82 Starburst}
\author[Fenech {\it et al.}]{D.\,M.\,Fenech,$\!^{1,2}$\thanks{Email:\,dmf@star.ucl.ac.uk} 
T.\,W.\,B.\,Muxlow$^2$, R.\,J.\,Beswick$^2$, A.\,Pedlar$^2$ and M.\,K.\,Argo$^3$\\
%\newauthor  %% starts a new line in the author environment
$^1$Department of Physics and Astronomy, University College London, Gower Street, London, WC1E~6BT\\
$^2$Jodrell Bank Centre for Astrophysics, School of Physics \& Astronomy, Alan Turing Building, University of Manchester, Manchester, M13 9PL\\
$^3$Department of Applied Physics, Curtin University of Technology, GPO Box U1987, Perth, Western Australia 6845}

%\date{Accepted ?. Received ?. Original form }
\pagerange{\pageref{firstpage}--\pageref{lastpage}} \pubyear{2008}

\maketitle
\label{firstpage}

\begin{abstract}
{The results of an extremely deep, 8-day long observation of the central kpc of the nearby starburst galaxy M82 using MERLIN (Multi-Element Radio Linked Interferometer Network) at 5\,GHz are presented. The 17\,\mujybm, rms noise level in the naturally weighted image make it the most sensitive high resolution radio image of M82 made to date. Over 50 discrete sources are detected, the majority of which are supernova remnants, but with 13 identified as {\HII} regions. 
Sizes, flux densities and radio brightnesses are given for all of the detected sources, which are all well resolved with a majority showing shell or partial shell structures. Those sources within the sample which are supernova remnants have diameters ranging from 0.3 to 6.7\,pc, with a mean size of 2.9\,pc.

From a comparison with previous MERLIN 5\,GHz observations made in July 1992, which gives a 9.75 year timeline, it has been possible to measure the expansion velocities of ten of the more compact sources, eight of which have not been measured before. These derived expansion velocities range between 2200 and 10500\,\kms.}
\end{abstract}


\section{Introduction}\label{Int}

A galaxy undergoing a period of rapid star formation that cannot be maintained for its lifetime is known as a starburst galaxy. M82, which is one of the closest known starburst galaxies \citep[3.2\,Mpc,][]{burbidge64}, is considered to be the archetypal example. At FIR wavelengths, starburst galaxies emit strongly, the emission arising from dust, heated by a population of early-type stars associated with the star formation \citep{rieke80}. This large population of massive, rapidly evolving stars will produce an equally large number of supernovae. The central regions of the majority of starbursts suffer from high visual extinction, making optical detection of these supernovae or supernova remnants (SNR) difficult. However, radio emission, which is unaffected by extinction, allows the central regions of these galaxies to be probed. The detected radio emission consists of both non-thermal (synchrotron) emission, arising from the acceleration of particles to relativistic speeds via supernova explosions \citep[e.g.][]{condon92}, and thermal free-free emission from {\HII} regions.

Radio observations of M82 in 1975 \citep{kronberg75} revealed a number of compact sources in the central kpc, which were considered to be supernovae. However, with the exception of the source, 41.95+57.5, whose flux density continues to decay (see section \ref{notes}), subsequent observations have shown a lack of variability in the flux densities of these sources which rule out the possibility of the majority of them being new radio supernovae and confirmed their identity as supernova remnants (SNR). Subsequent radio studies of the central region of M82 with MERLIN (Multi Element Radio Linked Interferometry Network) and the VLA (Very Large Array) have resulted in the detection of over 50 discrete objects which are now known to represent a population of both SNR and {\HII} regions \citep[e.g.][and others]{muxlow94,wills97,allen99,mcdonald02,rodr04}.

Studies of extragalactic SNRs have a number of advantages over similar Galactic ones. Firstly by the fact that the youngest known Galactic SNR, Cassiopeia A, is over 300 years old, (whereas the youngest known SNR in M82 is effectively 40 years old \citep{beswick06}). Secondly, the distances to Galactic SNRs are uncertain, but, although the relative distances of the SNRs in M82 differ by $\sim$ few hundred parsecs, that is small compared with the overall distance of 3.2\,Mpc so that they can effectively be treated as being at the same distance, thus providing a sample of supernova remnants that can be studied with essentially the same linear resolution and sensitivity. 
Global VLBI observations, which have previously been used to monitor the expansion of two of the most compact sources in M82 \citep[e.g.][]{pedlar99,mcdonald01,beswick06} resolve out all but the most compact sources. Hence, MERLIN is the ideal instrument for observations of this galaxy and this paper describes a MERLIN deep integration at 5\,GHz which provides high signal-to-noise ratios for the majority of the well resolved SNRs in M82. The naturally weighted sensitivity of $\sim$17\,\mujybm enables the study of larger, more diffuse, and by inference older, remnants amongst the SNR and {\HII} regions within the central kpc all of which are resolved in these 5\,GHz MERLIN observations. 

In \textsection\,\ref{msre}, measurements of the expansions of ten SNR from this sample, including eight for which this has not been done before, are presented. The resulting velocities range from 2200 to 10500\,\kms. In \textsection\,\ref{disc} the population of sources observed as well as the cumulative distribution of the SNRs and the flux density vs. diameter relation are discussed. These results, along with the star formation rates are used to investigate the discrepancy between the theoretical models of the SNRs and interstellar medium within M82 (\textsection\,\ref{ISM}), for example, the work of \cite{chevalier01}, and the results of the observations presented here.

\section{Observations and Image Processing\\}\label{OI}

\indent The 2002 deep integration was obtained using six of the MERLIN antennas, including the Cambridge telescope. The observing strategy and correlator configuration replicated those used in the 1992 5\,GHz observations presented by \cite{muxlow94}. The 2002 observations were undertaken between the 1 and 28 April at a frequency of 4.994\,GHz using both LL and RR polarisations and with a bandwidth of 15\,MHz divided into 15 channels. The total on source integration time was over 175 hours. Observations of the calibrator 3C286 were used to set the flux density scale and the point source calibrator OQ208 was used to determine the passbands and relative gains of the antennas. A phase reference source, 0955+697, was used to determine the telescope phases and the data were weighted appropriately according to the relative sensitivity of each MERLIN antenna. After applying phase corrections, the 2002 dataset was processed directly in the J2000 coordinate system using an updated position for the phase calibrator source, 0955+697, derived from VLBI observations.

In order for a direct comparison to be made, the 1992 dataset was precessed to J2000 coordinates and corrections were made to account for the revised position of the phase calibrator source. Final astrometrical alignment of the two datasets was achieved by aligning the peak of the most compact source (41.95+57.5) as measured in the 1992 image with that in a preliminary 2002 image. The peaks of the other compact sources (e.g. 44.01+59.6 and 43.31+59.2) were now found to be coincident to within 1\,mas. 

Following the alignment process both sets of data were imaged using  the AIPS task IMAGR to produce four adjoining 1024$\times$1024 fields, initially using natural weighting and a cell-size of 15\,mas. The images were deconvolved using the H\"{o}gbom cleaning algorithm \citep{hogbom74} and restored with a 50\,mas circular beam. The rms noise level in source free areas in the 2002 image was $\sim$17\,$\mu$Jy/beam, which was close to the expected noise level and a factor of $\sim$\,3 improvement on the $\sim46\,\mu$Jy/beam equivalent noise level in the 1992 images.

Subsequent images from both datasets were also produced using a variety of robustness parameters \citep{briggs95} and in some cases smoothing. The robustness values ranged from $-$3 to +3 with corresponding beam sizes ranging from 35 to 50\,mas. A summary of the imaging parameters used is given in Table \ref{params}. A uniformly weighted image was also produced and restored with a 35\,mas circular beam. As individual sources within M82 vary dramatically in nature with some being very bright and compact, whilst others are weak and more extended, the imaging parameters were adjusted to obtain the most suitable image for each source.

\section{Results}\label{Res}

\begin{landscape}
\begin{figure}
\centering
\includegraphics[width=16cm, angle=270]{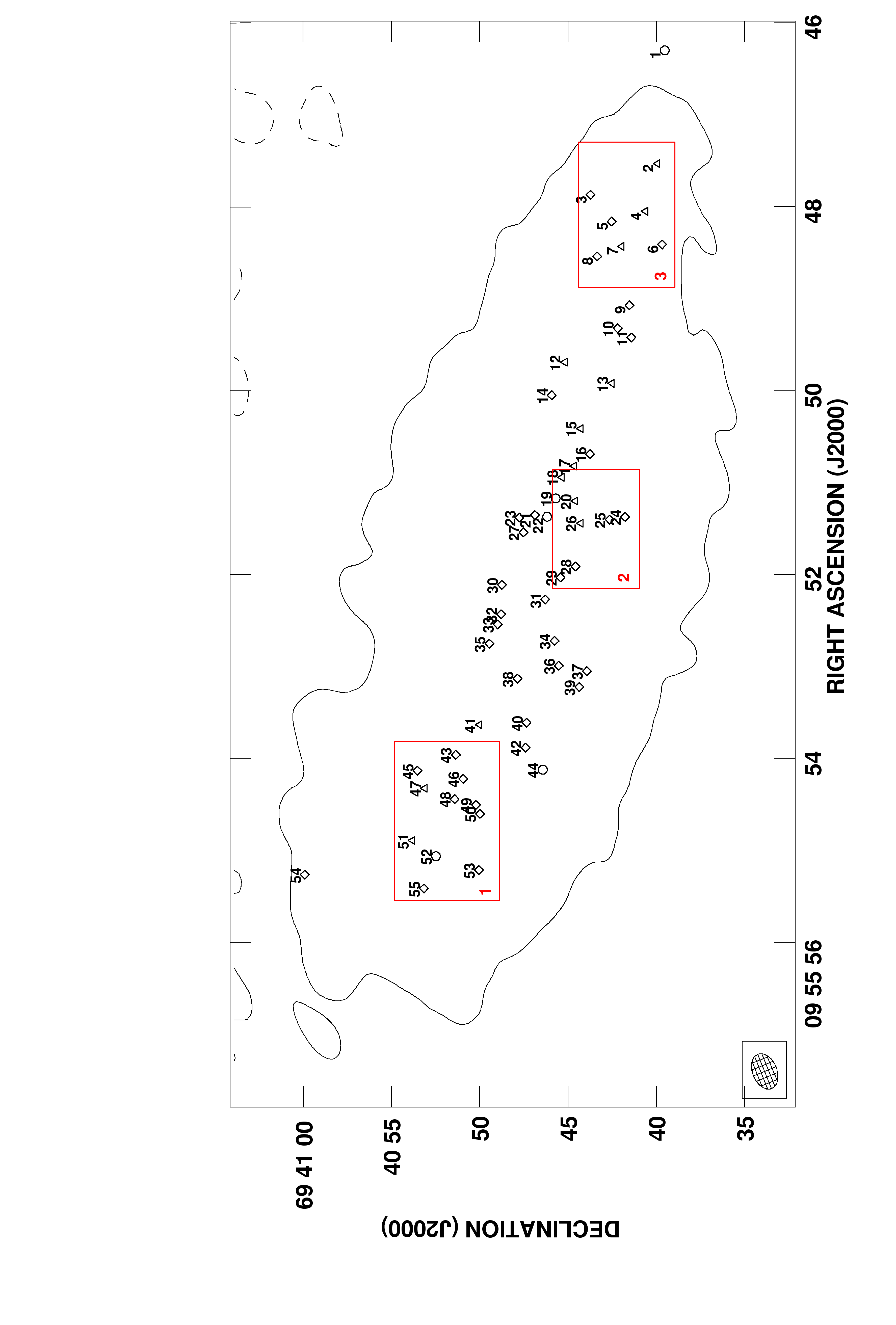}
\caption{The discrete sources discussed in this paper are identified in this image as; SNR (diamonds), {\HII} regions (triangles), unknown (circles). The rectangles show the areas covered by the contour and grey-scale images from the 2002 deep MERLIN observations presented in Fig. \protect{\ref{field}}. The single contour shown is the 1.2\,mJy\,$\rm{beam^{-1}}$ contour of an image of Very Large Array (VLA) A-array C-band observations of M82 from 2005 \protect{\citep{argo06}}.}
\label{m82full}
\end{figure}
\end{landscape}

\begin{figure}
\begin{center}
\includegraphics[width=23cm,angle=270]{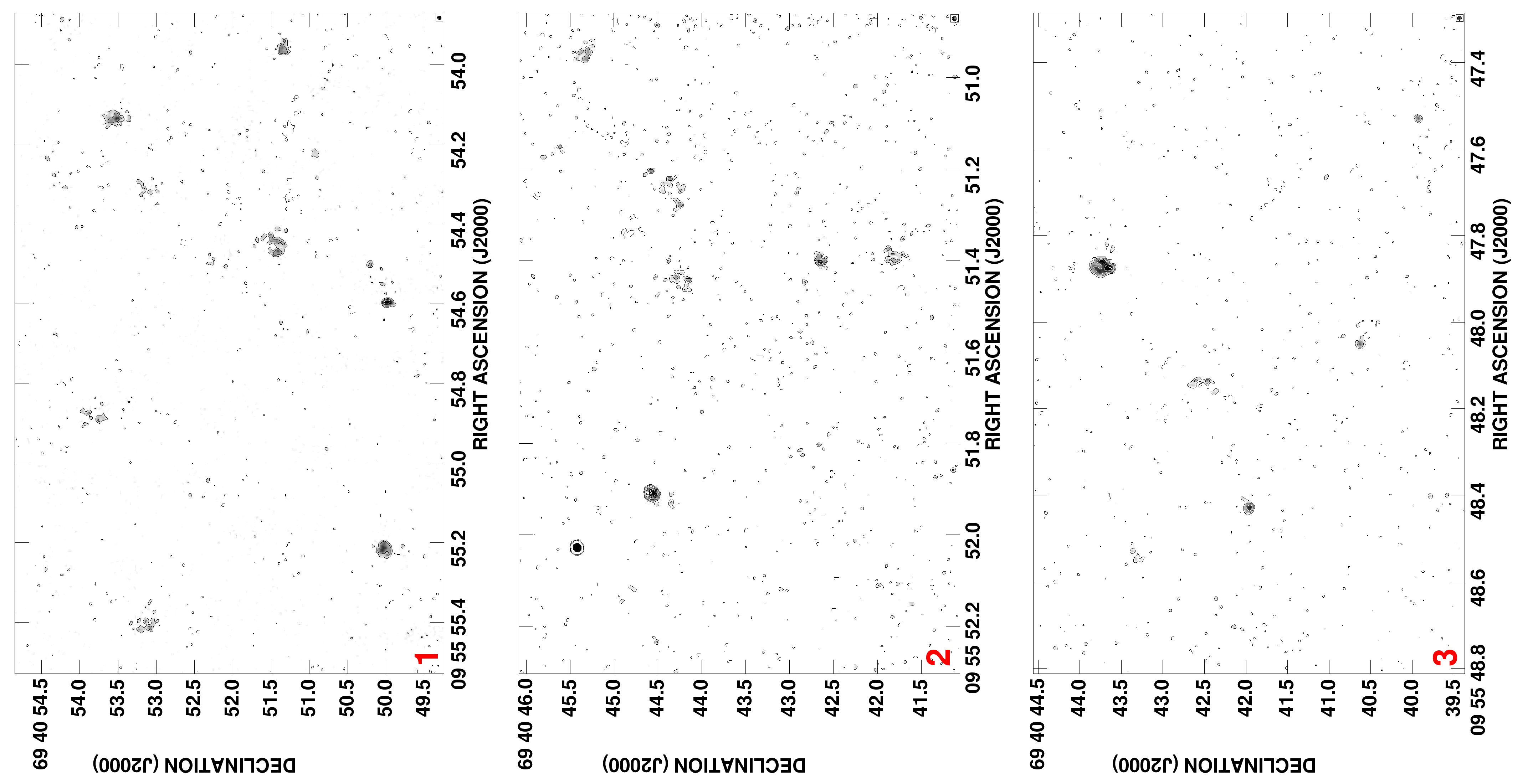}
\caption{Contour and grey-scale images of the naturally weighted maps of the 2002 deep observations convolved with a circular 50\,mas beam. The greyscale ranges from 51\,\mujybm to 300\,\mujybm and the contours are plotted at $-1,1,2,4,6,8,10,12,14,16,18,20\,\times\,$51\,\mujybm.}
\label{field}
\end{center}
\end{figure}

\begin{figure}
\begin{center}
\includegraphics[width=16cm]{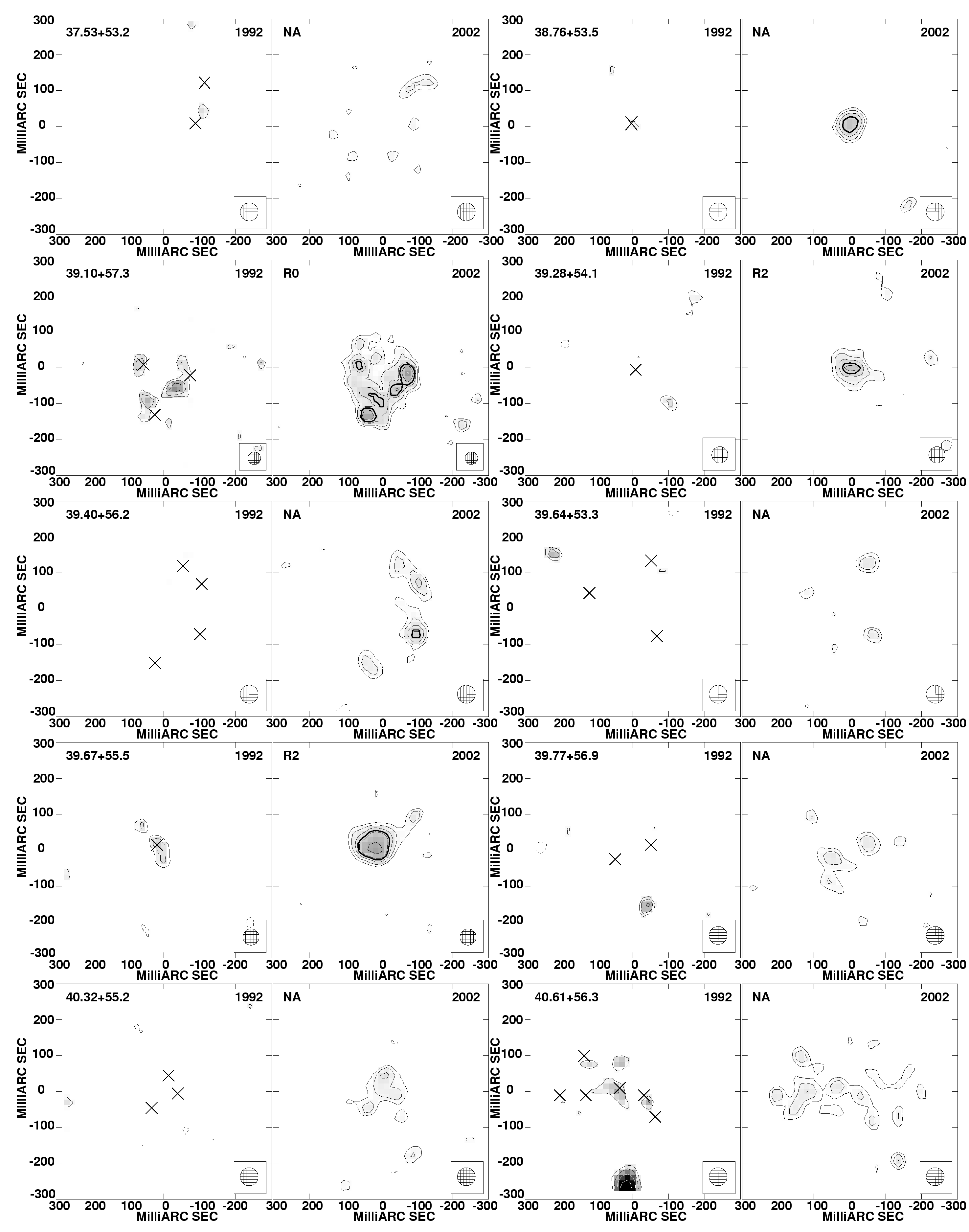}
\end{center}
\caption{Grey-scale and contour images of the individual sources as seen in the 1992 and 2002 observations. The grey-scale ranges from 130\,\mujybm to 600\,\mujybm for the 1992 data and from 49\,\mujybm to 600\,\mujybm for the 2002 data. The contours are plotted at $-1,\,1,\,1.414,\,2,\,2.828,\,4,\,5.656,\,8,\,11.282\,\times\, 3\rm{\sigma}$ noise level (see Table \ref{params}). The beam size is shown in the bottom right of each image and is listed in Table \ref{params}. The crosses in the 1992 images indicate the positions of the peaks of emission as observed in the 2002 images. The bold contour in the 2002 data represents the equivalent 3$\sigma$ contour at the 1992 noise level, though it should be noted that this will not be visible in all images as this may correspond to a brightness level above that of the source shown.} 
\label{contours}
\end{figure}

Fig. \ref{m82full} shows the 1.2\,mJy\,$\rm{beam^{-1}}$ contour in a VLA image of M82 obtained from observations in 2005 \citep{argo06}. Also shown are the positions of the compact sources identified in this paper. The rectangular boxes outline three areas, the images of which, from the 2002 MERLIN observations presented here, are shown in Fig. \ref{field} . Several shell or partial shell structures are visible in these images, which illustrate both the range in radio morphologies and relative sizes of the observed sources; these can be seen in more detail in the individual images in Fig. \ref{contours}.

Only sources with a peak flux density of $\geq$\,85\mujybm\,\,($\sim$5$\sigma$, $\rm{T_{B}> 1000\,K}$, 50\,mas) in the 2002 epoch have been included in Fig. \ref{contours}. This provides a sample of 55 sources, all of which are resolved by the 35\,mas beam. The positions, flux densities and deconvolved sizes of the detected sources are listed in Table \ref{tabA1}. The integrated flux densities were measured over areas that approximately match the shape of the 3$\sigma$ contour in the naturally weighted image. The mean background level in the vicinity of each source was also measured to correct the flux density values.

Fig. \ref{contours} shows the contour and greyscale images of each of the sources for both the 1992 and 2002 epochs using the same CLEANing parameters for both epochs ( i.e. the most appropriate for the 2002 data). Each of the contour plots is labelled, the label and the corresponding image parameters are listed in Table \ref{params}. Each of the contour levels are plotted as multiples of the 3$\rm{\sigma}$ noise levels, listed in Table \ref{params}, for both the 1992 and 2002 images. 
The naming convention of \cite{kronberg85}, identifying each source by its B1950 right ascension secs and declination arcsecs, has been used.

\begin{table}
\begin{center}
\caption{Parameters for the images used to create the contour and grey-scale plots of the sources presented in this paper. Rn represents the Robustness parameter used, where n is an integer from $-5$ to $+5$, S represents smoothing of the image and NA stands for natural weighting.}
\begin{tabular}{|c|c|c|c|}
\hline
\multicolumn{1}{c}{Robustness Value} & 
\multicolumn{1}{c}{Beam Size} & 
\multicolumn{2}{c|}{Rms noise ($\rm{\mu}Jybeam^{-1}$)} \\ %\cline{4-5}\\
 & (mas) & 2002 & 1992 \\
\hline 
 R-1 & 35 & 24 & 60\\
 R0 & 35 & 22 & 52\\
 R0S & 45 & 23 & 54\\	
 R1 & 41 & 20 & 49\\
 R1S & 48 & 21  & 50 \\
 R2 & 45 & 19 & 48\\
 R3 & 46 & 18 & 47\\
 NA & 50 & 17 & 46\\
\hline
\end{tabular}
\label{params}
\end{center}
\end{table}

%\clearpage

\clearpage
%\twocolumn
\subsection{Measuring diameters}\label{sizes}

The radio sources detected in these observations can be broadly split into three categories; very compact sources; diffuse shells or partial shells and extended low surface brightness sources. A number of methods have been used to measure the diameters of the detected sources dependent upon this variety of observed radio morphologies. The diameters of the very compact sources (single peak, $\lesssim\,100$\,mas in diameter) were measured using a two dimensional Gaussian fit to determine the full width half maximum of the source. The more diffuse, easily identifiable shell or partial-shell, structured sources (flux density $\,\gtrsim\,0.650\,\rm{mJy}$) were measured using a combination of integrated annular profiles and Gaussian fitting to the discrete knots within the source. For the extended, weak sources or those showing no obvious shell structure, flux density profiles obtained by drawing slices across the source were used to measure the source sizes.

The name, position, peak, integrated flux density, and deconvolved size for each source are listed in Table \ref{tabA1} together with a source identification where possible. For those sources whose structures are approximately circular, an average diameter is given. Otherwise major and minor axis measurements are listed. For the weaker sources (peak flux $\lesssim\,0.1\,\rm{mJy\,beam^{-1}}$) a largest angular size (LAS) is quoted. The distance taken for M82 of 3.2\,Mpc provides a linear size equivalent to 1\,milliarcsecond\,$\equiv$\,0.0155\,pc. For the purposes of measuring diameters, the {\HII} regions have been treated in the same manner as the SNR, although, the two classes will be discussed individually in the following section.

\newpage
%\onecolumn
\newpage

\begin{figure}
\begin{center}
\includegraphics[width=16cm]{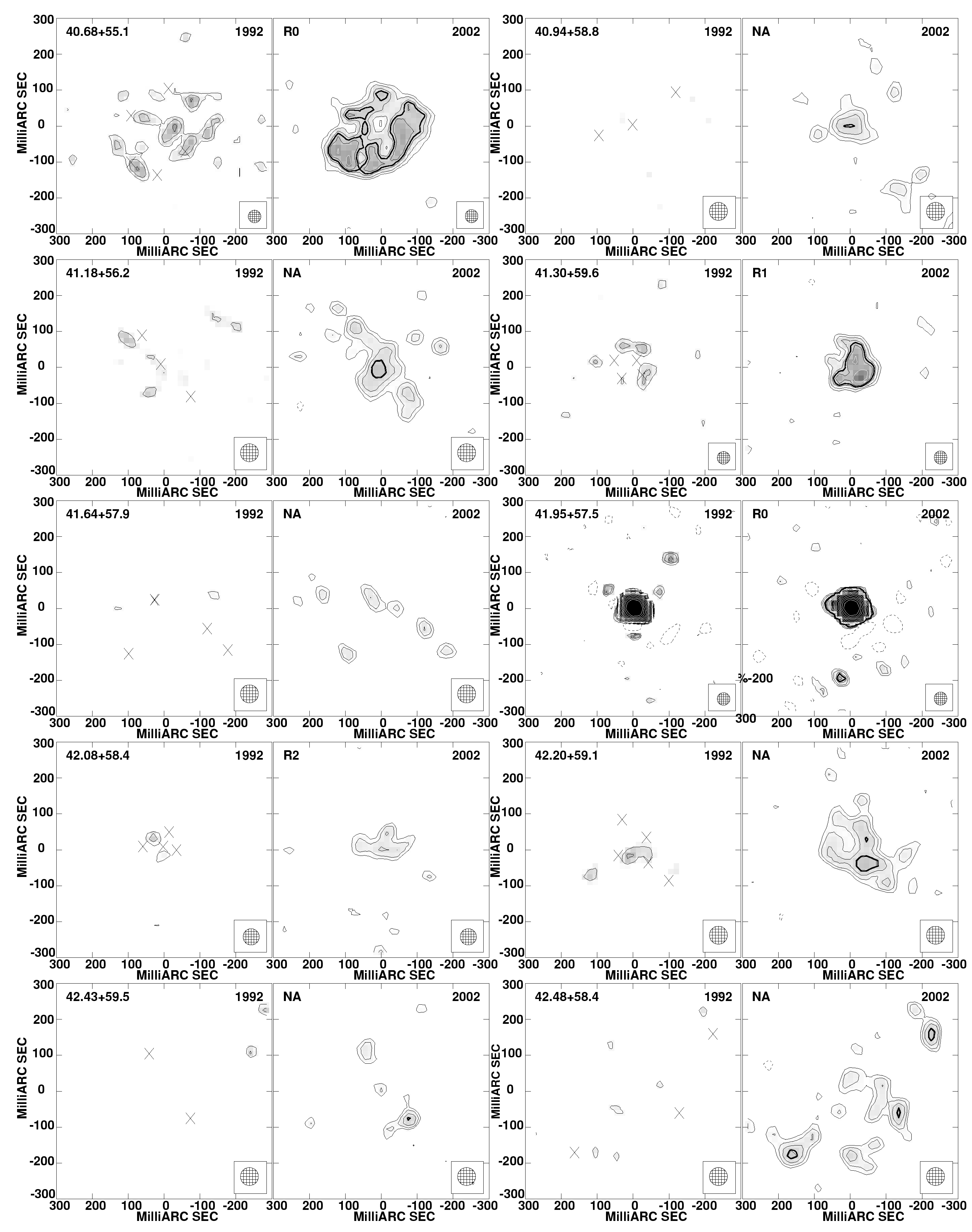}
\hfill\\
\it Figure 3 Continued 
\end{center}
\end{figure}
\newpage

\begin{figure}
\begin{center}
\includegraphics[width=16cm]{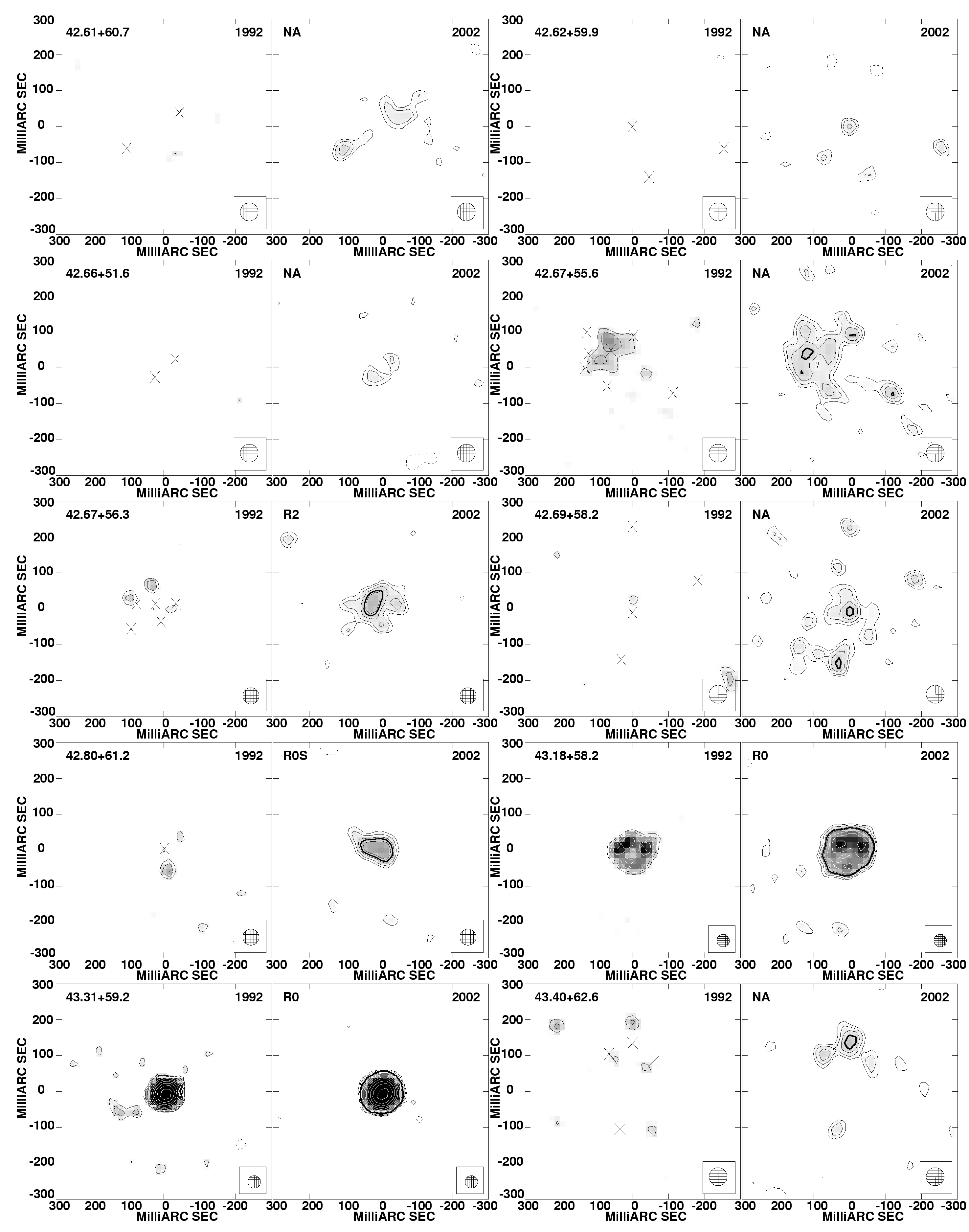}
\hfill\\
\it Figure 3 Continued 
\end{center}
\end{figure}
\newpage

\begin{figure}
\begin{center}
\includegraphics[width=16cm]{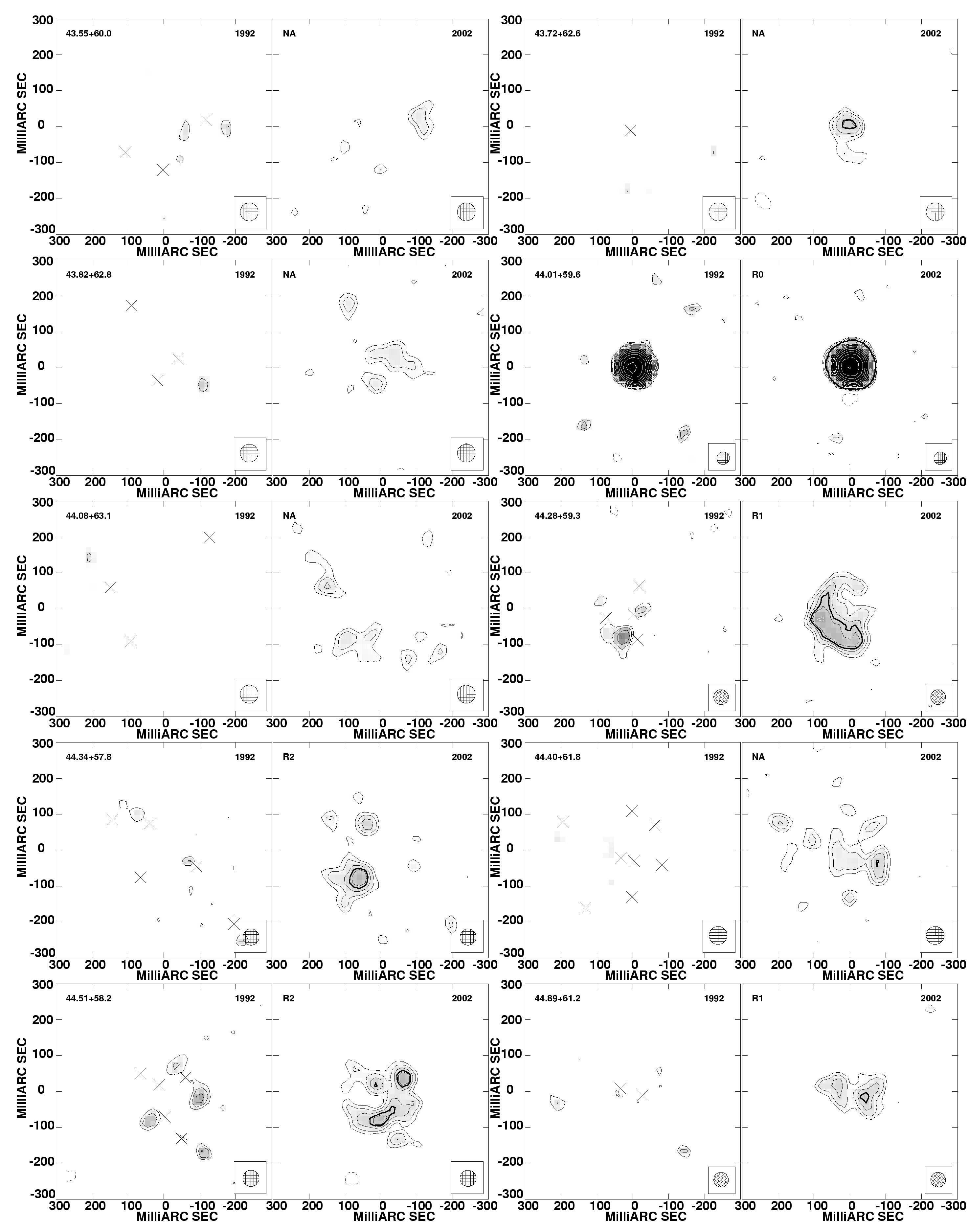}
\hfill\\
\it Figure 3 Continued 
\end{center}
\end{figure}
\newpage

\begin{figure}
\begin{center}
\includegraphics[width=16cm]{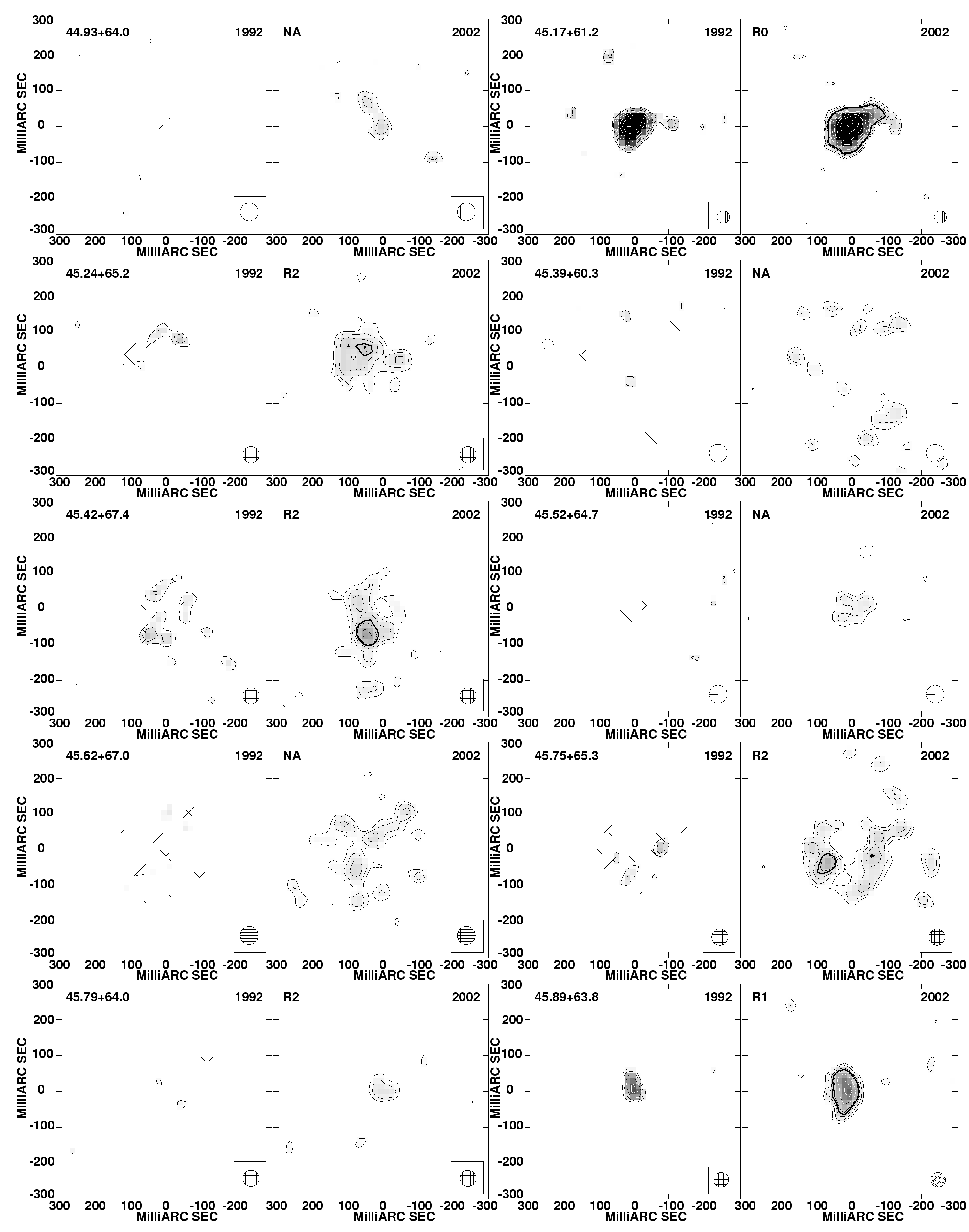}
\hfill\\
\it Figure 3 Continued 
\end{center}
\end{figure}
\newpage

\begin{figure}
\begin{center}
\includegraphics[width=16cm]{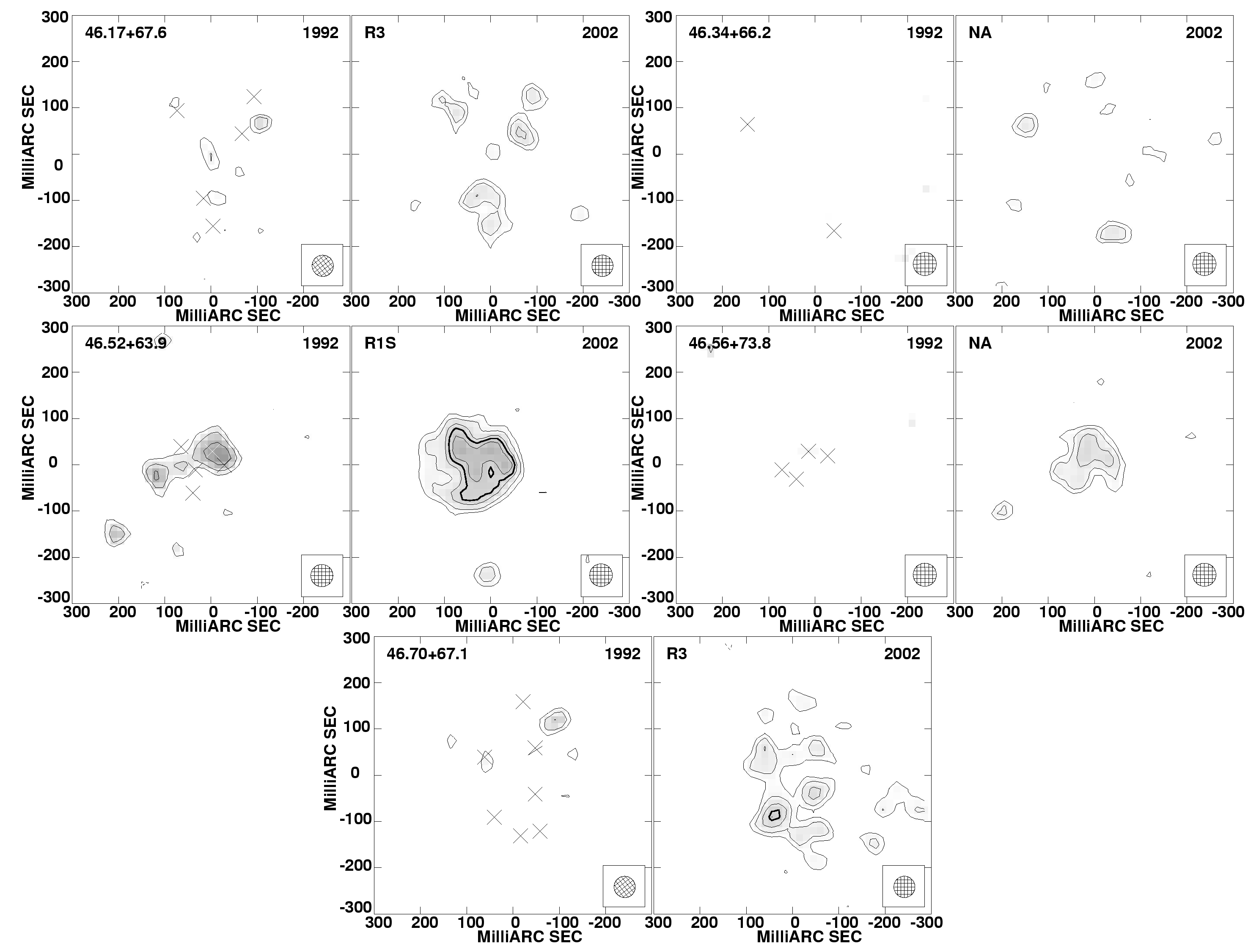}
\hfill\\
\it Figure 3 Continued 
\end{center}
\end{figure}

\newpage
%\onecolumn

\renewcommand{\thefootnote}{\fnsymbol{footnote}}

\begin{small}
\begin{center}
\begin{longtable}[h]{lllcccccl}
\caption{Flux densities and deconvolved sizes of the sources detected in the 2002 5-GHz MERLIN dataset. Those sources previously identified in the literature as SNRs or {\HII} regions have been labelled in the table. These identifications are taken from \protect\cite{muxlow94,wills97,allen99,mcdonald02}. The right ascension and declination are offset from $\rm{09^{h}\,55^{m}\,00^{s}}$ and $+69^{o}\,40'\,00''$ (J2000). Errors are shown in brackets where appropriate.}
\label{tabA1}
\tabularnewline 

\hline
  \multicolumn{1}{l}{Number} &
  \multicolumn{1}{c}{Name} &
  \multicolumn{1}{c}{RA} &
  \multicolumn{1}{c}{Dec} &
  \multicolumn{1}{c}{Peak Flux} &
  \multicolumn{1}{c}{Integrated Flux} &
  \multicolumn{2}{c}{Size} &
  \multicolumn{1}{c}{Comments}\\
& & & & Density & Density & & &	\\
& & J2000 & J2000 &  (mJy/beam) & (mJy) & mas & pc &\\
\hline
\hline
\endfirsthead

\multicolumn{9}{l}
{{\it \tablename\ \thetable{} -- continued from previous page}}\\ 

\hline
  \multicolumn{1}{l}{Number} & 
  \multicolumn{1}{c}{Name} &
  \multicolumn{1}{c}{RA} &
  \multicolumn{1}{c}{Dec} &
  \multicolumn{1}{c}{Peak Flux} &
  \multicolumn{1}{c}{Integrated Flux} &
  \multicolumn{2}{c}{Size} &
  \multicolumn{1}{c}{Comments}\\ 
& & & & Density & Density & & &	\\
& & J2000 & J2000 &  (mJy/beam) & (mJy) & mas & pc &\\
\hline	
\hline 
\endhead
\multicolumn{9}{l}{{\it Continued on next page}} \endfoot
\endlastfoot
\hline
1 & 37.53+53.2 & 46.30 & 39.52 & 0.085 (0.016) & 0.121 (0.024) & 475.1 & 7.4 & Unknown\footnote{Sources labelled as unknown have had no prior identification as either SNR or {\HII} region, as there is little or no spectral index information available. These sources will be discussed individually in section \protect\ref{disc}.}\\
2 & 38.76+53.5 & 47.53 & 39.93 & 0.181 (0.017) & 0.272 (0.030) & 54.8(7.1)$\times$36.5(5.9) & 0.9$\times$0.6 & {\HII}\\
3 & 39.10+57.3 & 47.87 & 43.73 & 0.381 (0.019) & 3.555 (0.084) & 219(10) & 3.4 & SNR\\% Shell\\
4 & 39.28+54.1 & 48.05 & 40.62 & 0.197 (0.016) & 0.468 (0.041) & 113.2(14.1)$\times$67.9(9.4) & 1.8$\times$1.0 & {\HII}\\
5 & 39.40+56.2 & 48.16 & 42.53 & 0.154 (0.018) & 0.820 (0.071) & 361 & 5.6 & SNR\\% Partial Shell\\
 & & & & & & & &\\
6 & 39.64+53.3 & 48.41 & 39.68 & 0.099 (0.016) & 0.099 (0.019) & 241 & 3.7 & SNR\\
7 & 39.67+55.5 & 48.43 & 41.96 & 0.283 (0.017) & 0.808 (0.042) & 107.2(8.1)$\times$85.1(6.7) & 1.7$\times$1.3 & {\HII}\\
8 & 39.77+56.9 & 48.54 & 43.36 & 0.089 (0.016) & 0.180 (0.022) & 262\,LAS & 4.1 & SNR\\
9 & 40.32+55.2 & 49.07 & 41.53 & 0.114 (0.017) & 0.253 (0.030) & 76 & 1.2 & SNR \\
10 & 40.61+56.3 & 49.32 & 42.19 & 0.104 (0.016) & 0.409 (0.039) & 429\,LAS & 6.7 & SNR \\%Partial\\
 & & & & & & & &\\
11 &  40.68+55.1 & 49.42 & 41.42 & 0.484 (0.016) & 6.093 (0.080) & 255 & 4.0 & SNR \\%Shell\\
12 & 40.94+58.8 & 49.69 & 45.19 & 0.142 (0.020) & 0.264 (0.039) & 266$\times$75 & 4.1$\times$1.2 & {\HII}\\
13 & 41.18+56.2 & 49.92 & 42.51 & 0.177 (0.019) & 0.830 (0.056) & 262$\times$122 & 4.1$\times$1.9 & {\HII}\\  
14 & 41.30+59.6 & 50.05 & 45.92 & 0.565 (0.027) & 2.296 (0.081) & 126 & 2.0 & SNR \\%Partial\\
15 & 41.64+57.9 & 50.41 & 44.29 & 0.105 (0.017) & 0.236 (0.035) & 421 LAS & 6.5 & {\HII}\\
 & & & & & & & &\\
16 & 41.95+57.5 & 50.69 & 43.76 & 14.395 (0.028) & 17.035 (0.073) & 23.6(0.1)$\times$14.3(0.2) & 0.4$\times$0.2 & SNR? Bi-polar\footnote{The true nature of this source and its identification as a SNR is uncertain. This will be discussed further in section \protect\ref{notes}}\\
17 & 42.08+58.4 & 50.82 & 44.67 & 0.125 (0.018) & 0.321 (0.037) & 165$\times$73 & 2.6$\times$1.1 & {\HII}\\
18 & 42.20+59.1 & 50.94 & 45.34 & 0.178 (0.018) & 1.038 (0.061) & 172 & 2.7 & {\HII}\\
19 & 42.43+59.5 & 51.17 & 45.70 & 0.095 (0.015) & 0.179 (0.023) & 261 & 4.0 & Unknown\\
20 & 42.48+58.4 & 51.20 & 44.59 & 0.160 (0.018) & 0.973 (0.067) & 572$\times$202 & 8.9$\times$3.0 & {\HII}?\\
 & & & & & & & &\\  
21 & 42.61+60.7 & 51.35 & 46.89 & 0.127 (0.019) & 0.300 (0.040) & 254$\times$168 & 3.9$\times$2.5 & SNR \\
22 & 42.62+59.9 & 51.37 & 46.19 & 0.117 (0.019) & 0.238 (0.036) & 283 & 4.4 & Unknown\\
23 & 42.66+51.6 & 51.38 & 47.76 & 0.091 (0.018) & 0.205 (0.041) & 80 & 1.2 & SNR\\ 
24 & 42.67+55.6 & 51.37 & 41.78 & 0.161 (0.024) & 1.104 (0.082) & 215 & 3.3 & SNR \\%Partial\\
25 & 42.67+56.3 & 51.40 & 42.65 & 0.239 (0.021) & 0.843 (0.063) & 86 & 1.3 & SNR \\%Partial\\
 & & & & & & & &\\
26 & 42.69+58.2 & 51.44 & 44.29 & 0.157 (0.018) & 0.545 (0.057) & 375 & 5.8 & {\HII}\\
27 & 42.80+61.2 & 51.54 & 47.53 & 0.263 (0.019) & 0.545 (0.044) & 93.6(8.1)$\times$51.6(4.9) & 1.5$\times$0.8 & SNR \\%Shell\\
28 & 43.18+58.2 & 51.91 & 44.58 & 0.943 (0.027) & 3.908 (0.088) & 125.3 & 1.9 & SNR \\%Shell\\
29 & 43.31+59.2 & 52.03 & 45.42 & 5.273 (0.018) & 9.530 (0.051) & 48.2(0.31)$\times$40.1(0.27) & 0.7$\times$0.6 & SNR \\%Shell\\
30 & 43.40+62.6 & 52.11 & 48.74 & 0.174 (0.024) & 0.454 (0.059) & 276$\times$166 & 4.3$\times$2.6 & SNR \\%Shell\\
 & & & & & & &\\
31 & 43.55+60.0 & 52.27 & 46.31 & 0.098 (0.017) & 0.131 (0.024) & 273\,LAS & 4.2 & SNR\\
32 & 43.72+62.6 & 52.43 & 48.79 & 0.154 (0.017) & 0.274 (0.030) & 98.1(15)$\times$55.9(10) & 1.5$\times$0.9 & SNR\\
33 & 43.82+62.8 & 52.54 & 48.98 & 0.101 (0.016) & 0.322 (0.035) & 283\,LAS & 4.4 & SNR \\%Partial\\
34 & 44.01+59.6 & 52.72 & 45.77 & 9.984 (0.024) & 19.435 (0.072) & 51.3(0.2)$\times$49.4(0.2) & 0.8$\times$0.8 & SNR \\%Partial\\
35 & 44.08+63.1 & 52.75 & 49.46 & 0.112 (0.017) & 0.400 (0.040) & 403 & 6.2 & SNR \\%Partial\\
 & & & & & & & &\\
36 & 44.28+59.3 & 52.99 & 45.53 & 0.335 (0.019) & 1.642 (0.063) & 206 & 3.2 & SNR \\%Partial\\
37 & 44.34+57.8 & 53.05 & 43.94 & 0.242 (0.019) & 0.954 (0.061) & 147 & 2.3 & SNR\\
38 & 44.40+61.8 & 53.13 & 47.87 & 0.142 (0.023) & 0.631 (0.063) & 288 & 4.5 & SNR\\
39 & 44.51+58.2 & 53.22 & 44.36 & 0.212 (0.018) & 1.446 (0.067) & 178 & 2.8 & SNR \\%Shell\\
40 & 44.89+61.2 & 53.61 & 47.35 & 0.212 (0.017) & 0.858 (0.049) & 130$\times$63 & 2.0$\times$1.0 & SNR\\
 & & & & & & & &\\
41 & 44.93+64.0 & 53.63 & 50.01 & 0.093 (0.017) & 0.211 (0.034) & 127\,LAS & 2.0 & {\HII}\\
42 & 45.17+61.2 & 53.88 & 47.41 & 2.427 (0.021) & 6.310 (0.069) & 75.9(1.1)$\times$47.4(1.0) & 1.2$\times$0.7 & SNR \\%Partial\\
43 & 45.24+65.2 & 53.96 & 51.36 & 0.193 (0.019) & 1.065 (0.063) & 219 & 3.4 & SNR \\%Partial\\
44 & 45.39+60.3 & 54.12 & 46.43 & 0.099 (0.021) & 0.447 (0.055) & 520\,LAS & 8.1 & Unknown\\
45 & 45.42+67.4 & 54.13 & 53.53 & 0.271 (0.016) & 1.152 (0.059) & 202 & 3.1 & SNR\\% Partial\\
 & & & & & & & &\\
46 & 45.52+64.7 & 54.22 & 50.94 & 0.095 (0.017) & 0.173 (0.029) & 98\,LAS & 1.5 & SNR\\% Partial\\
47 & 45.62+67.0 & 54.32 & 53.12 & 0.122 (0.016) & 0.853 (0.061) & 200 & 3.1 & {\HII}\\
48 & 45.75+65.3 & 54.44 & 51.43 & 0.228 (0.019) & 1.522 (0.087) & 251 & 3.9 & SNR \\%Shell\\
49 & 45.79+64.0 & 54.50 & 50.22 & 0.121 (0.015) & 0.192 (0.025) & 188$\times$61 & 2.9$\times$0.9 & SNR\\
50 & 45.89+63.8 & 54.60 & 49.98 & 0.482 (0.016) & 1.240 (0.044) & 88.6(4)$\times$46.3(2) & 1.3$\times$0.7 & SNR\\
 & & & & & & & &\\
51 & 46.17+67.6 & 54.89 & 53.81 & 0.128 (0.017) & 1.149 (0.081) & 213 & 3.3 & {\HII}\\
52 & 46.34+66.2 & 55.06 & 52.48 & 0.101 (0.017) & 0.207 (0.026) & 350\,LAS & 5.4 & Unknown\\
53 & 46.52+63.9 & 55.21 & 50.05 & 0.130 (0.018) & 0.650 (0.068) & 172 & 2.7 & SNR \\%Shell\\
54 & 46.56+73.8 & 55.26 & 59.90 & 0.130 (0.018) & 0.556 (0.054) & 124 & 1.9 & SNR \\%Partial\\
55 & 46.70+67.1 & 55.41 & 53.16 & 0.159 (0.018) & 1.440 (0.097) & 254$\times$140 & 3.9$\times$2.2 & SNR\\% Partial\\
\hline
\end{longtable}
\end{center}
\end{small}

\subsection{Notes on particular sources}\label{notes}

\paragraph*{41.95+57.5} The most compact source in M82, 41.95+57.5, has shown a continued decrease in flux density of $\sim$8.5\% per year since its first observation in 1965 \citep{trotman96}. At an age of $\sim$80 years (calculated assuming a free expansion of $\sim$2000\,\kms and a diameter of 0.3\,pc), the above decay rate would imply the source to have had a flux density of $\sim$30\,Jy at birth. It has been extensively studied using both EVN and MERLIN measurements. The most recent VLBI observations \citep{beswick06} show 41.95+57.5 to be bi-polar in nature, which is consistent with our results from the two-dimensional fitting giving a major axis of 23.6\,mas and a minor axis of 14.3\,mas with a position angle of 53.5$^{o}$. 
 Its continued flux density decay and the implication that at its birth it would have been a remarkably bright radio source suggests that 41.95+57.5 could be the remains of a gamma-ray burst (GRB) event rather than a SNR. This would also provide an explanation for the double-lobed nature of the source as seen in the global VLBI observations \citep{beswick06}. This has been discussed in greater detail in \cite{muxlow05}.
\paragraph*{41.30+59.6} This SNR has shown a large increase in flux density (from 1.4 to 2.3\,mJy) over the 9.75 years between two MERLIN 5-GHz observations. The extent of this increase, coupled with the fact that (with the exception of 41.95+57.5), no other source shows such dramatic behaviour, can rule out the possibilty of this being the result of a calibration error. Its internal structure has also evolved significantly between the two epochs which is not unusual for such sources. Several other sources (such as 40.68+55.1 and 41.95+57.5) have also shown strong structural evolution.
\paragraph*{40.59+55.8} Although this source is clearly visible (with a flux density of 1.23\,mJy and a size of 1.2\,pc) at the bottom of the field showing the image of the source, 40.61+56.3, in the 1992 observations (see Fig. \ref{contours}), it is not detected in the new 2002 dataset with a detection limit of 21\,$\rm{\mu}Jy$, and was undetected in MERLIN multi-frequency synthesis 5\,GHz observations in 1999 \citep{mcdonald02}. Therefore, it has been classified as a transient source, together with the source, 41.50+59.7, detected by \cite{kronberg85} which has a similar detection limit of $\sim$20\,$\rm{\mu}Jy$ in the current 5-GHz observations.

\section{Monitoring supernova remnant expansions}
\label{msre}

Where possible, the same methods used to measure source sizes using the 2002 data have been applied to the re-imaged 1992 dataset so that direct comparisons can be made. The alignment of the two datasets to within 1\,mas at a distance of 16.5 arcseconds from the phase centre has enabled the measurement of expansion velocities for a number of sources. Whilst all of the observed motion cannot necessarily be explained by expansion, it is most likely that the majority of the observed radial changes will be a consequence of expansion. The results are presented in Table \ref{tab2}. This has only been possible for the stronger sources (higher flux density) because of the lower signal-to-noise ratio of the original 1992 dataset. However, for many of these, this has been a first opportunity to measure their expansion as all but the most compact sources have been totally resolved in previous VLBI observations \citep[e.g.][]{pedlar99, mcdonald01, beswick06}.
All of the sources for which this has been done are supernova remnants. Whilst it is likely that the {\HII} regions will show some expansion, it will be difficult to detect with typical expansion velocities of $\sim$10-100\,\kms. No {\HII} regions were detected in the 1992 dataset because of the lower signal-to-noise.
 
\subsection{Compact supernova remnants}\label{cexp}
\begin{figure}
\begin{center}
\includegraphics[width=5cm,angle=270]{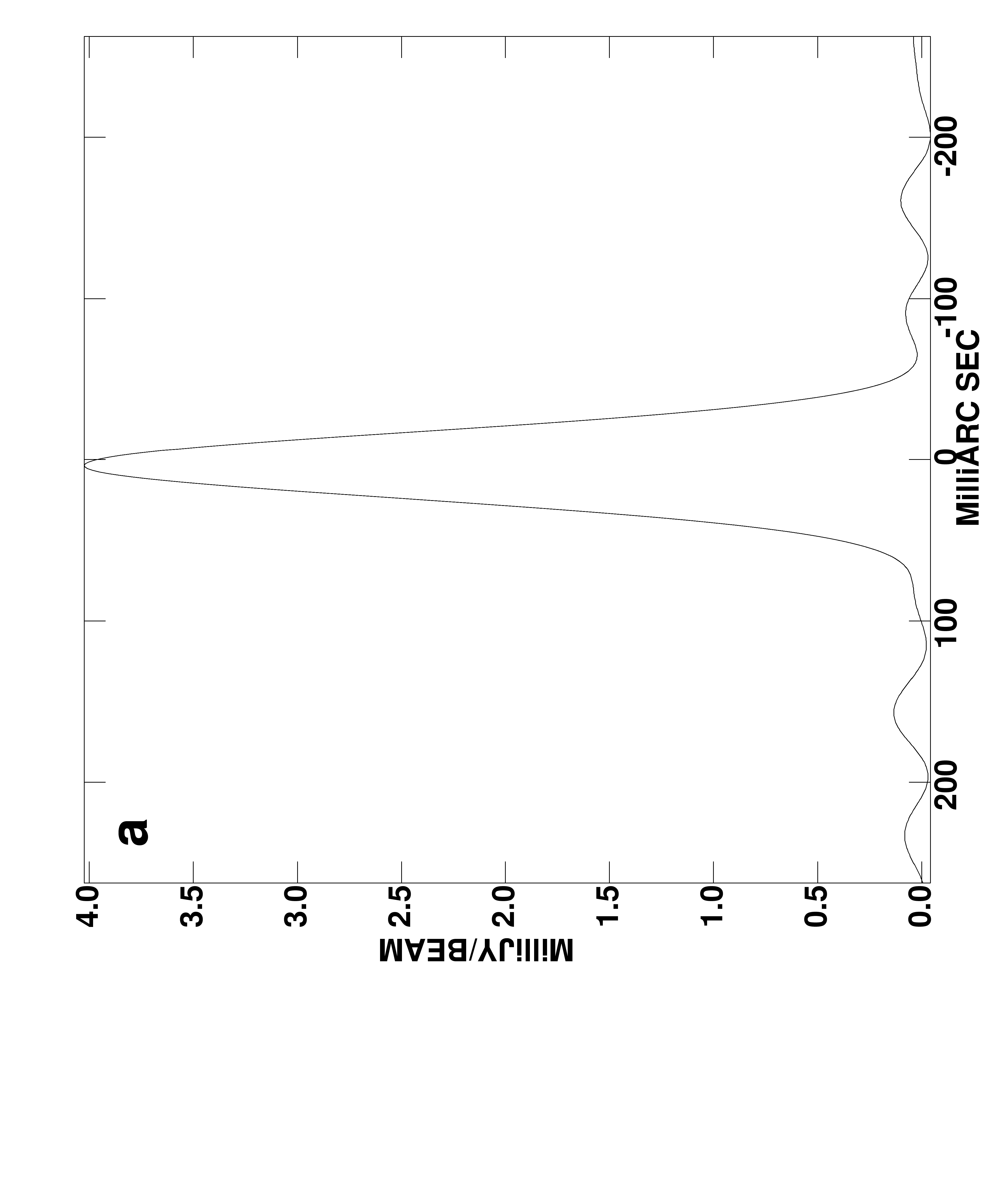}
\includegraphics[width=5cm,angle=270]{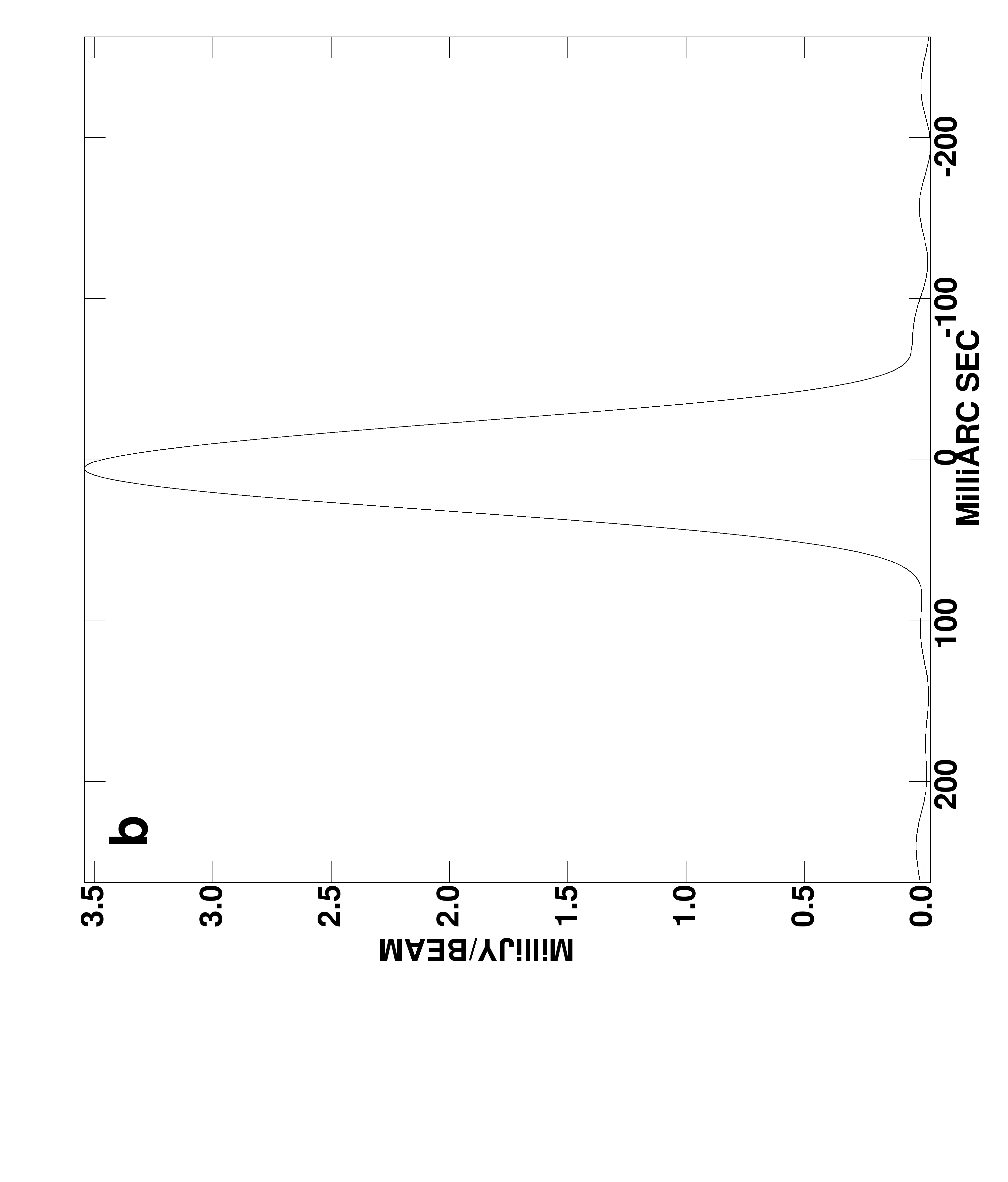}
\includegraphics[width=5cm,angle=270]{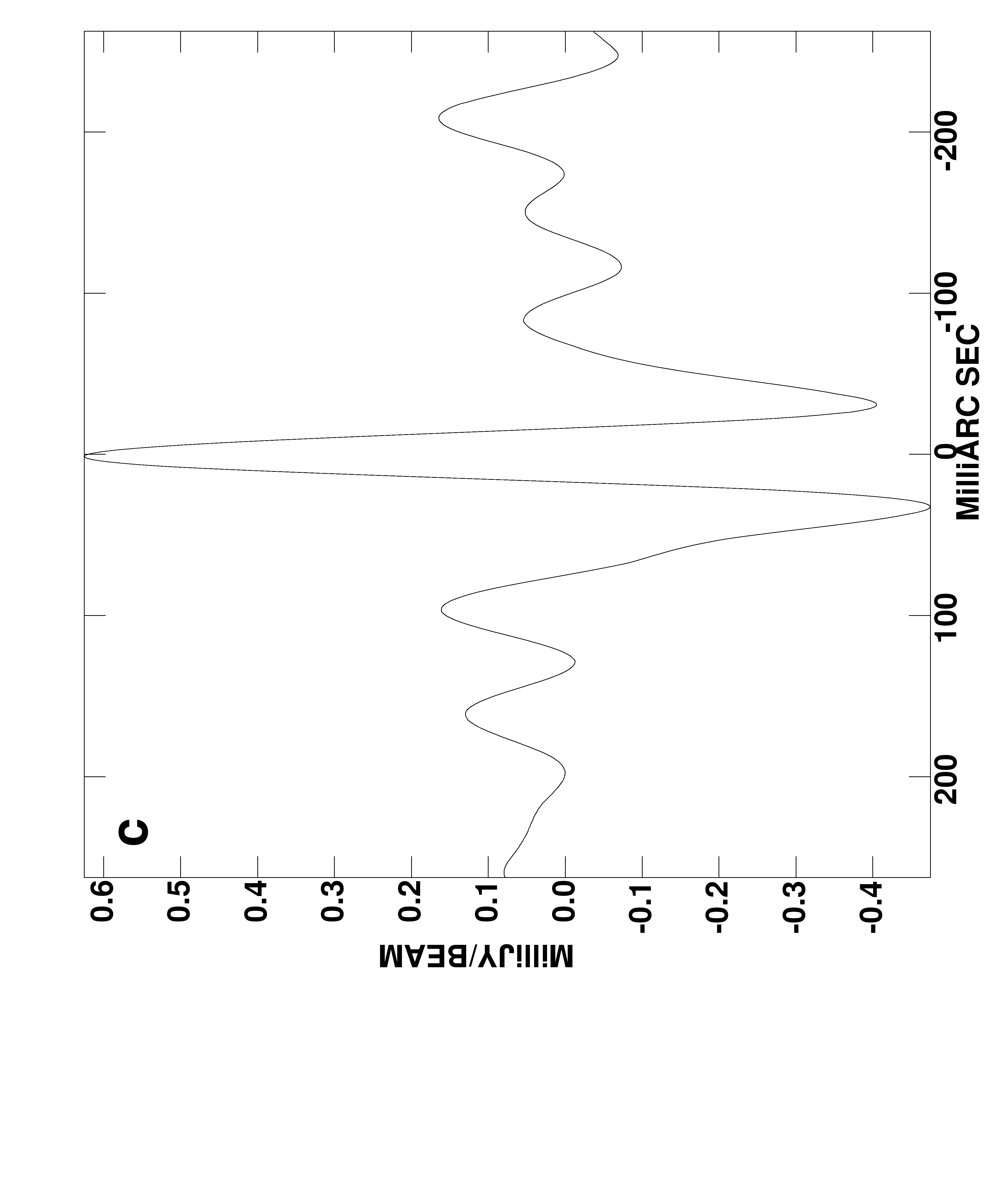}
\caption{Slices across the centre of the source 43.31+59.2, from the 1992 (a) and 2002 (b) dataset, as well as across the subtracted image (c) (the 2002 image subtracted from the 1992 image) which shows the expected inner peak and surrounding negative region.}
\label{exp}
\end{center}
\end{figure}

\paragraph*{41.95+57.5} For the marginally resolved sources, Gaussian fitting to the full width half maximum has been used to measure the size and expansion velocity of the source. In the case of 41.95+57.5, the expansion velocity has been calculated along the major axis 
because of the unusual structure of this source (see Section \ref{notes}). It shows an increase in size of 2.88$\pm0.13$\,mas corresponding to a velocity of 2200$\pm$200\,\kms. This is in good agreement with measurements from VLBI observations which give values of 1500-2000\,\kms \citep{pedlar99, mcdonald01, beswick06}. 
\paragraph*{43.31+59.2} This source has been studied in detail using VLBI observations over the last 15 years at 18\,cm. \cite{beswick06} measure its expansion to be 9025$\pm$380\,\kms from the most recent VLBI measurements and 11000$\pm$1000\,\kms using a Common Point Method \citep{marcaide06}. From the Gaussian fitting to the two epochs of data we find this source to be expanding with a velocity of 8800$\pm$600\kms (using an average size), again in agreement with the VLBI measurements. Fig. \ref{exp} shows a flux density slice drawn across the residual image of 43.31+59.2 after subtracting the 2002 image from the 1992 image. This clearly illustrates the observed expansion of this remnant between the two epochs showing the expected inner peak surrounded by a negative outer region.
\paragraph*{44.01+59.6} Though seen as a compact, relatively unresolved source in these MERLIN observations, the expansion of 44.01+59.6 has been difficult to measure using VLBI observations, a consequence of its relatively large size at VLBI resolutions. These MERLIN observations have provided one of the first opportunities to reliably measure the expansion of this source, which is found to be 2700$\pm$400\,\kms.

\subsection{Shell-like remnants}\label{sexp}

The expansion velocities for the shell-like sources have been measured using integrated annular profiles as well as Gaussian fitting to the peaks of emission, where possible. The central pixel of the source was estimated by eye. Annular profiles were then calculated, centred on each pixel of a 5$\times$5 grid, itself centred on the estimated source centre. The GNUPLOT package was then used to fit a two-component Gaussian to the resulting central annular profile, to estimate the peak of the shell and hence calculate the expansion velocity. This method is preferable for estimating the expansion of the larger, more diffuse SNR, as it does not rely on the correct identification of individual features over multiple epochs.

\begin{figure}
\begin{center}
\includegraphics[width=5.6cm,angle=270]{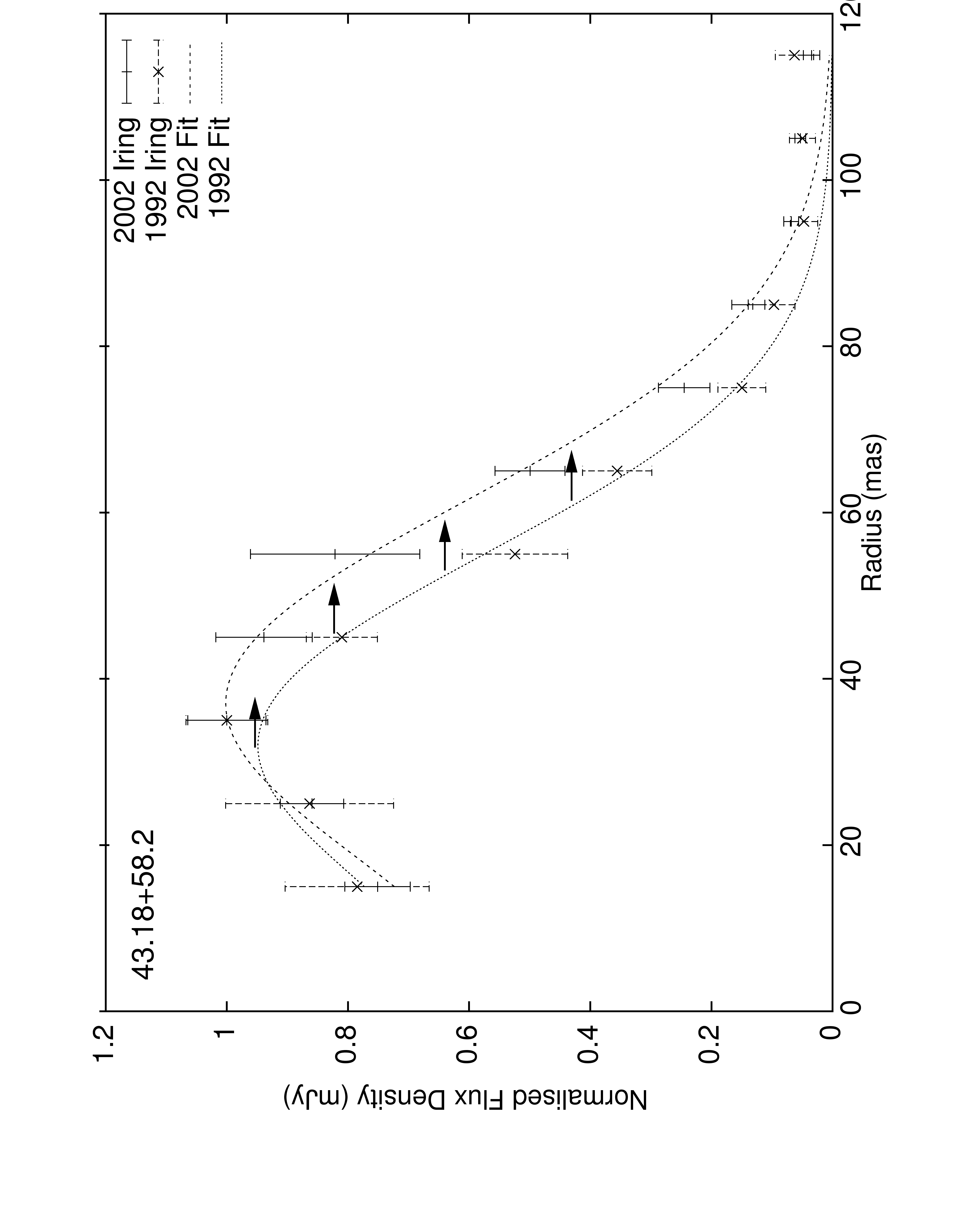}
\includegraphics[width=5.8cm,angle=270]{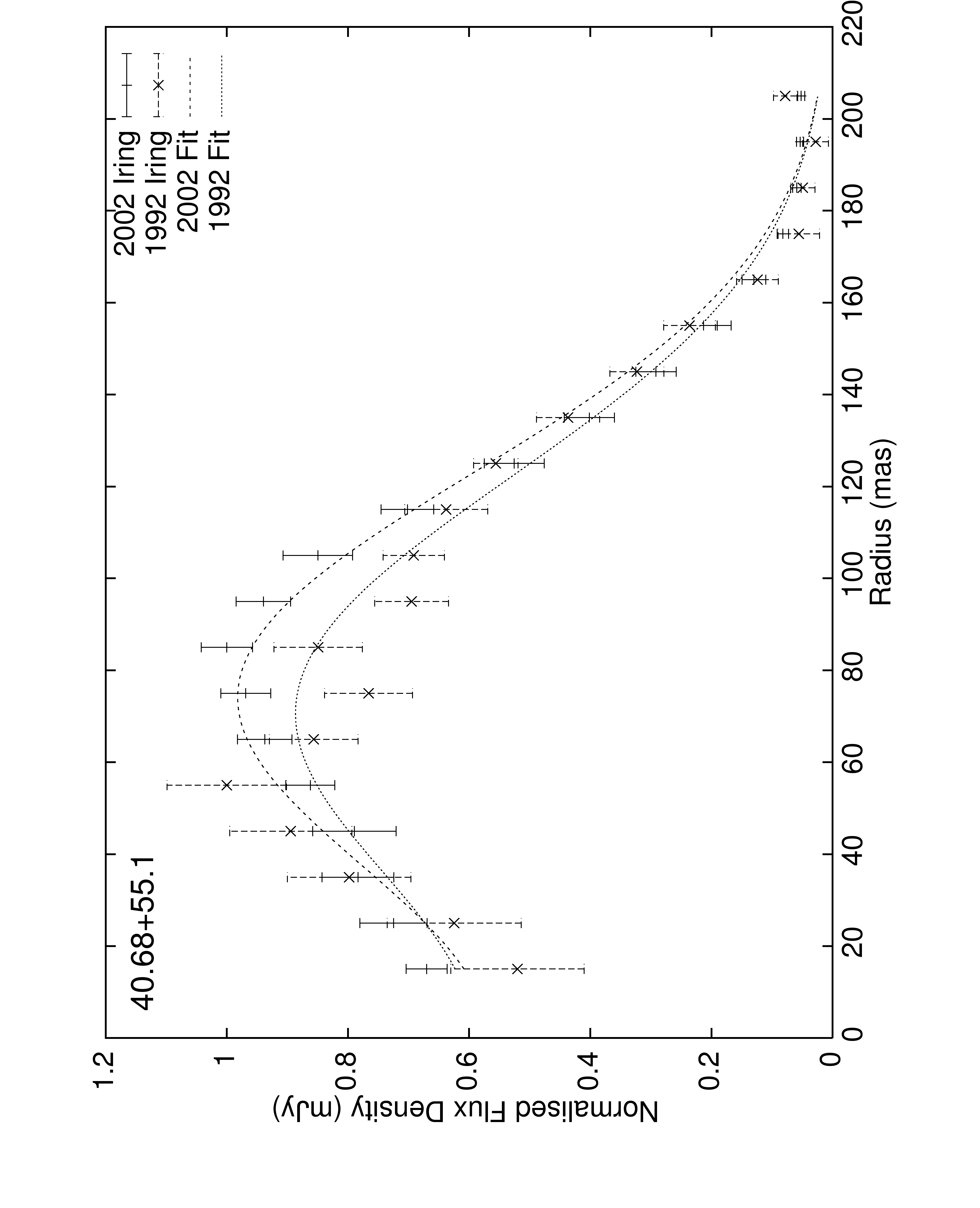}
\caption{An example of the annular profiles for two of the supernova remnants. Each image shows the integrated annular profiles for both epochs. A fit to the profile data using the GNUPLOT package is shown and used to estimate the expansion velocity.} 
\label{exp2}
\end{center}
\end{figure}

\begin{figure}
\begin{center}
\includegraphics[width=12.7cm,angle=0]{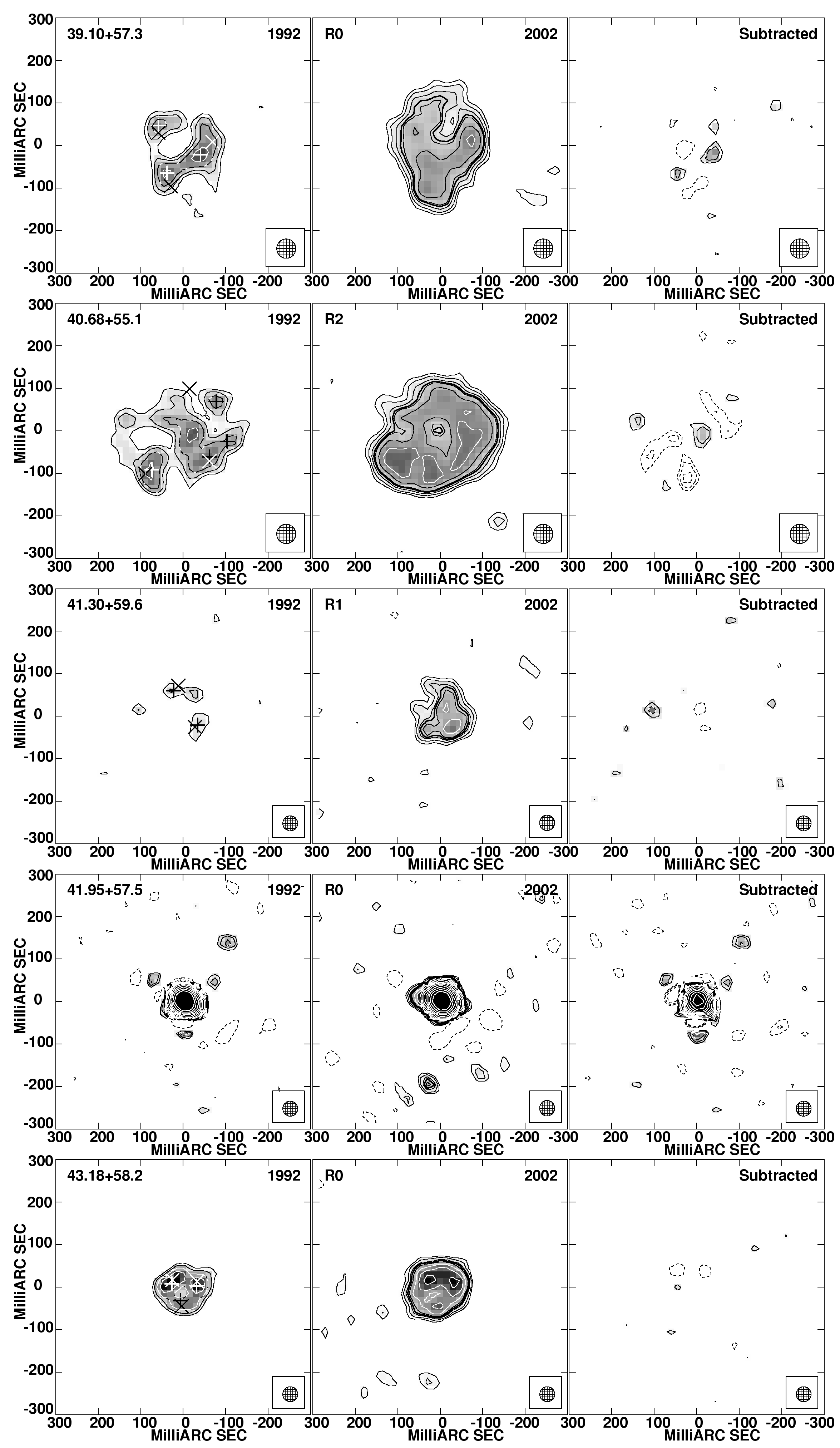}
\caption{Contour and grey-scale plots of the sources for which expansions have been measured. Left: 1992 image, middle: 2002 image, right: residual image after the 2002 image has been subtracted from the 1992 image. Contours are plotted at $-1,\,1,\,1.414,\,2,\,2.828,\,4,\,5.656,\,8,\,11.282,\,16,\,22.564,\,32\,\times\, 3\rm{\sigma}$ (see Table \protect\ref{params}) and the grey-scale ranges from 49 to 600\,\mujybm for the 2002 images and 130 to 600\,\mujybm for the 1992 and subtracted images. The subtracted images have additional contours plotted at $\rm{-2\,and -1.414 \times 3\,\sigma}$. The bold contour in the 2002 images represents the equivalent 3$\sigma$ contour at the 1992 noise level. The crosses and pluses in the 1992 image show the 2002 and 1992 peak positions respectively.}
\label{expfig}
\end{center}
\end{figure}

\paragraph*{43.18+58.2} The SNR 43.18+58.2 (Fig. \ref{expfig}) shows a clear, almost complete, shell structure and Gaussian fits to the peaks of the radio emission within the shell have shown them to have moved outward by 7.0$\pm$0.3\,mas in the 9.75 years between observations, providing an expansion velocity of 10500$\pm$900\,\kms. This is confirmed by a comparison of the integrated annular profiles for this source showing the radius to have increased from 30.4 to 37.2\,mas, as can be seen in Fig. \ref{exp2}, giving a velocity of 10500$\pm$2300\,\kms. Whilst not unusually high for a sample of SNRs, this velocity is the highest of those measured, and given its proximity to 43.31+59.2 (which also has a high expansion velocity) might suggest that both of these sources are embedded in a low-density region of the ISM, a possibility discussed by \cite{beswick06} for the case of 43.31+59.2. However, no low frequency turnover has been detected for 43.31+59.2 (see section \ref{disc}), whereas \cite{wills97} show the emission from 43.18+58.2 to have a distinct low frequency turnover. This suggests that 43.31+59.2 and 43.18+58.2 may occupy a very different location in the starburst region, with 43.18+58.2 possibly located deep within or even on the far side of the starburst region, which would be consistent with the observed significant free-free absorption along the line-of-sight to it.
\paragraph*{39.10+57.3} In the case of 39.10+57.3, there is a very distinct shell structure and an increase in the annular profiles from 64.3 to 69.0\,mas shows the source to be expanding at 7600$\pm$3400\,\kms. Gaussian fitting to the peaks of the emission confirm this expansion showing an increase in separation by 5.0\,mas. The motion of the peaks however, does not appear to be completely uniform (seen Fig. \ref{expfig}) which may partially be caused by the source expanding into an inhomogeneous environment. However, the shell structure seen in both epochs and the definite increase in radius show it to be expanding.

\paragraph*{40.68+55.1} For the SNR 40.68+55.1 and 44.28+59.3, the integrated annular profiles provide the best evidence for the observed expansion. 40.68+55.1 has a definite shell structure and is observed with good signal to noise in both epochs. The internal structure of this source has changed significantly over the 9.75 years; hence monitoring the expansion via fitting to individual peaks of emission is extremely difficult. However, a comparison of the annular profiles for both epochs (see Fig. \ref{exp2}) shows an increase in peak radius from 145.2 to 150.4\,mas giving an expansion velocity of 4000$\pm$3400\,\kms.
\paragraph*{44.28+59.3} Measurements of the expansion for the majority of the sources has not been possible because of the low sensitivity of the original 1992 dataset. The SNR, 44.28+59.3, shows an example of a source \textbf{very} close to this limiting factor. The partial shell-structure evident in the images from both epochs has enabled a tentative measurement of its expansion velocity of $\sim$ 3700\,\kms. This has been measured using a comparison of the integrated annular profiles, and is included as an illustrative example of the limitation imposed by the low sensitivity of the 1992 dataset. 
\paragraph*{45.17+61.2 and 45.89+56.3} The sources 45.17+61.2 and 45.89+56.3 could both be interperated as either compact SNRs or discrete radio knots within a larger shell. Assuming the latter is correct, velocities from the shifts in positions of the peaks of emission between the two epochs have been calculated. In the case of 45.17+61.2 the expansion velocity is $\sim$6000\,\kms, although it does not appear to be radial if the source is believed to be a partial shell structure. Consequently, the true nature of the motion of this source is unclear.

The source 45.89+56.3 also shows a large shift in the peak position between the two epochs potentially indicating an expansion of $\sim$6100\,\kms. \\

\paragraph*{41.30+59.6} As discussed in section \ref{notes}, 41.30+59.6 is an unusual source. Its radio morphology has changed dramatically over the period between the two observations. Consequently, measuring the expansion velocity is not trivial. However, Gaussian fitting to the peaks of emission has been used to measure the change in diameter, providing a tentative expansion velocity of 5000$\pm$4500\,\kms.

\onecolumn

\begin{figure}
\begin{center}
\includegraphics[width=14cm,angle=0]{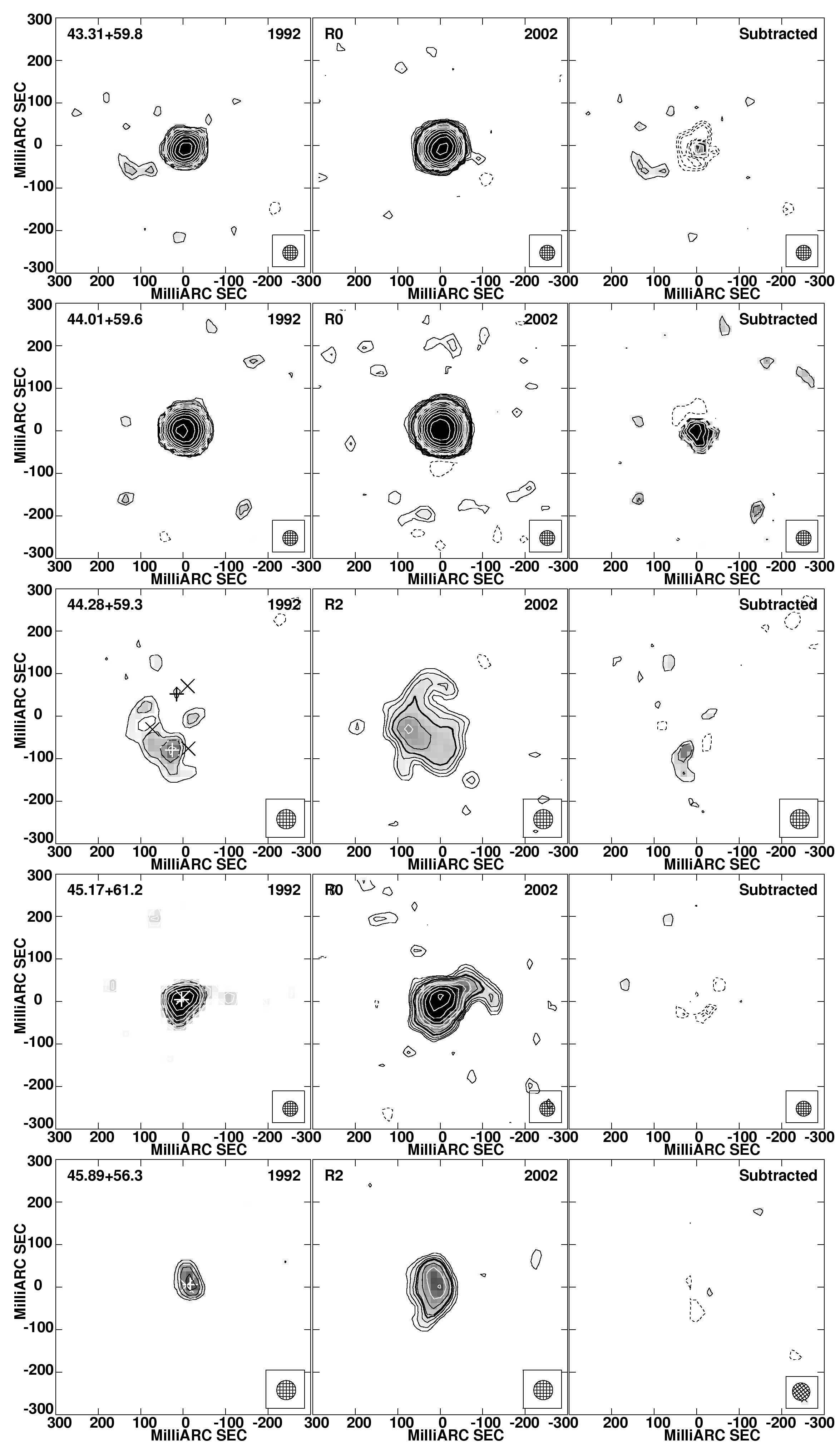}\\
\it Figure \ref{expfig} continued.
\end{center}
\end{figure}

%\onecolumn
\begin{small}
\begin{landscape}
\begin{table}
\begin{center}
\caption{The expansion velocities calculated for the SNR including the peak and integrated flux densities from the 1992 observations and the source sizes in both epochs used to calculate the expansion (with the exception of 45.17+51.2 and 45.89+63.8, whose expansions were calculated using a peak offset). The errors are listed in brackets where appropriate and for the sizes, represent either the errors from the best fit to the annular profiles or the errors calculated from the Gaussian fitting to peaks of emission. Also listed are the expansion velocities previously published for the most compact sources. \citep[{[1]}][]{beswick06}, \citep[{[2]}][]{mcdonald02phd}.}
\begin{tabular}{|l|c|c|c|c|c|c|c}
\hline
  \multicolumn{1}{|c|}{Name} &
  \multicolumn{1}{c|}{Peak Flux (1992)} &
  \multicolumn{1}{c|}{Flux Density (1992)} &
  \multicolumn{2}{c|}{Diameter (mas)} &
  \multicolumn{1}{c|}{Radial velocity} &	
  \multicolumn{1}{c|}{Age in Free} &
  \multicolumn{1}{c|}{Other Velocities}\\ 
  & (mJy/beam) & (mJy) & 1992 & 2002 & (\kms) & Expansion (yrs) &\\
\hline
  39.10+57.3 & 0.53 & 3.60\,(0.16) & 128.2\,(1.8)& 138.0\,(2.6)& 7600\,(3400) & 140 & -\\
  40.68+58.3 & 0.47 & 5.49\,(0.25) & 145.2\,(2.6) & 150.4\,(2.0)& 4000\,(3400) & 280  & -\\
  41.30+59.6 & 0.43 & 1.41\,(0.09) & 98.1\,(2.8)& 104.7\,(3.4)& 5000\,(4500) & 160 & -\\
  41.95+57.5 & 34.94 & 39.01\,(0.19) & 20.7\,(0.1)$\times$8.4\,(0.1)& 23.6\,(0.1)$\times$14.3\,(0.1)& 2200\,(200) & 81 & 1500-2000\,\kms\,$^{[1]}$\\
  43.18+58.2 & 1.06 & 3.85\,(0.14) & 60.8\,(1.8)& 74.4\,(1.2)& 10500\,(2300) & 55 & $<$3270\,\kms\,(1-$\rm{\sigma}$ limit)\,$^{[2]}$\\
  43.31+59.2 & 5.69 & 8.51\,(0.11) & 38.4\,(0.6)$\times$29.6\,(0.5) & 49.6\,(0.3)$\times$ 41.2\,(0.3)& 8800\,(600) & 39 & 9000-11000\,\kms\,$^{[1]}$\\
  44.01+59.6 & 11.29 & 21.13\,(0.20) & 47.8\,(0.3)$\times$46.3\,(0.3) & 51.3\,(0.2)$\times$49.4\,(0.2) & 2700\,(400) & 140 & 4900$\pm$500\,\kms\,$^{[2]}$\\
  44.28+59.3 & 0.45 & 1.72\,(0.24) & 118.8\,(5.4)& 123.6\,(2.2)& $\sim$ 3700 & 250 & -\\
  45.17+61.2 & 2.65 & 5.44\,(0.13) & 39.2\,(0.6)$\times$29.3\,(0.5)& 73.3\,(1)$\times$46.0\,(0.8)& $\sim$ 6000 & 75 & 8800$\pm$4900\,\kms\,$^{[2]}$\\
  45.89+63.8 & 0.52 & 0.72\,(0.12) & 72.1\,(7)$\times$45.6\,(5)& 88.6\,(4)$\times$46.3\,(2)& $\sim$ 6100 & 85 & -\\
\hline
\end{tabular}
\label{tab2}
\end{center}
\end{table}
\end{landscape}

\end{small}

%\twocolumn
\section{Discussion}\label{disc}

\begin{figure}
\begin{center}
\includegraphics[width=10cm]{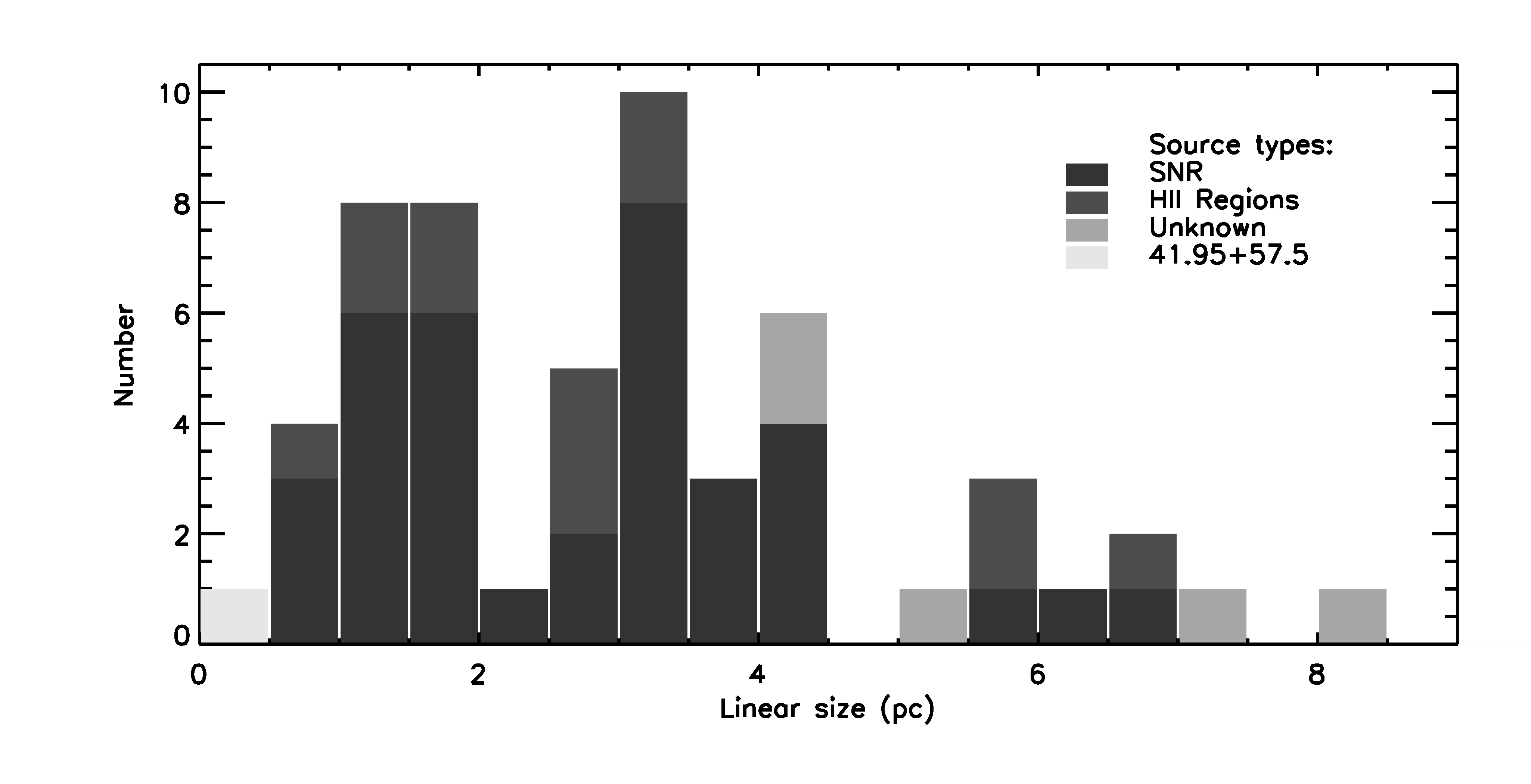}
\caption{Histogram showing the distribution of source sizes in the new 2002 observations, showing the SNR, {\HII} regions and the unidentified sources separately.}
\label{Hist}
\end{center}
\end{figure}

We have detected and imaged a sample of 55 of the discrete sources in M82, 13 of which are {\HII} regions and 37 are SNRs, identified by their steep spectra and high brightness temperatures, together with shell or partial shell structures. Five sources have no prior spectral identification, including the source, 42.43+59.5, which has been detected for the first time in these deep MERLIN observations with a flux density of $\sim$0.2\,mJy and a size of 4.0\,pc. 
The sources that show no obvious shell-like structure could be either regions of radio emission within a more extended low brightness shell or compact SNRs. Information for these sources has been included in spite of this ambiguity. Fig. \ref{Hist} shows a histogram of the size distribution of the sources observed, showing {\HII} regions, SNRs and unidentified sources separately.

\subsection{Unidentified sources}\label{unknown}

Five of the sources detected in this dataset have no prior identification as either an {\HII} region or SNR. In all cases, this is because of a lack of spectral index information. The source, 42.43+59.5, has been detected for the first time and 46.34+66.2 has only been detected at two wavelengths including 5\,GHz. Consequently, their identification is not possible at present and they will not be included in the SNR analysis. The sources, 37.53+53.2 and 45.39+60.3, have three point spectral energy distributions (SEDs), presented in \cite{allen99} from VLA observations of M82, which suggest that both these sources have a flat spectrum and are {\HII} regions. Their lack of detection in the VLA A-array 15-GHz data presented in \cite{mcdonald01} may be due to the fact that they are both large ($>7\,\rm{pc}$) and faint (peak flux density $<0.1\,\,\rm{mJy\,beam^{-1}}$). As such, these sources will also not be included in the SNR analysis in section \ref{ND}.\\
\indent The source, 42.62+59.9, also has a three-point SED presented in \cite{allen99}, indicating that it has a steep spectrum and is therefore probably a SNR and will be included in the analysis. Once again it has not been detected at 15-GHz observations by \cite{mcdonald01} probably because its flux density is below the noise level in the 15-GHz image.

\subsection{Sources of thermal origin.}\label{Hii}

Thirteen of the sources detected are believed to be {\HII} regions because of their flat spectra \citep{wills97,mcdonald02}. In M82 the {\HII} regions are thought to be associated with star formation \citep{condon92}, and HST observations have indicated that most of them seem to be associated with super-star clusters \cite[see][and others]{melo05,mcgrady07} in which OB stars provide the ionising radiation. The thermal sources observed cover a range in size from 0.7 to 6.5\,pc with a mean radius of 3.1\,pc (as illustrated in Fig. \ref{Hist}). The weakest {\HII} region detected has a flux density of 0.21\,mJy, corresponding to a brightness temperature of $\sim$1000\,K (127\,mas).

\indent The source referred to in this paper as 42.48+58.4 is comprised of the sources 42.48+58.4 and 42.56+58.0, identified by \cite{mcdonald02} as two individual radio knots. However, it can now be seen in Fig. \ref{contours} that there is considerable extended emission associated with these two sources, consequently they have been identified as part of a large {\HII} region. There is the possibilty that the new emission is from a separate source visible along the line of sight and appearing to be associated with the originally known {\HII} radio region. Future observations of appropriate sensitivity will ascertain whether this more diffuse emission also has a flat spectrum. It is interesting to note that this source is the largest observed in this sample with a major axis of 8.5\,pc.

\begin{table}
\begin{center}
\caption{A comparison of the sources detected in the two datasets showing shell-like structures.}
\begin{tabular}{|l|c|c}
\hline
 \multicolumn{1}{|l|}{Sources} &
 \multicolumn{2}{c|}{Number}\\
& 1992 & 2002 \\
\hline
SNR & 24 & 38 \\
{\HII} Regions & - & 13 \\
Unidentified & - & 4 \\
\hline
Total Detected & 24 & 55 \\
\hline
Shell/Partial Shell & 16 & 41 \\
\hline
\end{tabular}
\label{shells}
\end{center}
\end{table}

\subsection{Supernova remnants}\label{SNR}

A large proportion of the sample of sources, identified as supernova remnants, have steep spectra and shell or partial shell structures, many with what could be interpreted as outbreaks or gaps within the shell structure. A total of $\sim$41 sources show shell-like structures in the new dataset compared with $\sim$16 in the original 1992 dataset \citep{muxlow94}, a direct result of the achieved three-fold improvement in the noise level of the new data.

\begin{figure}
\begin{center}
\begin{minipage}[h]{8cm}
\includegraphics[width=8cm,angle=270]{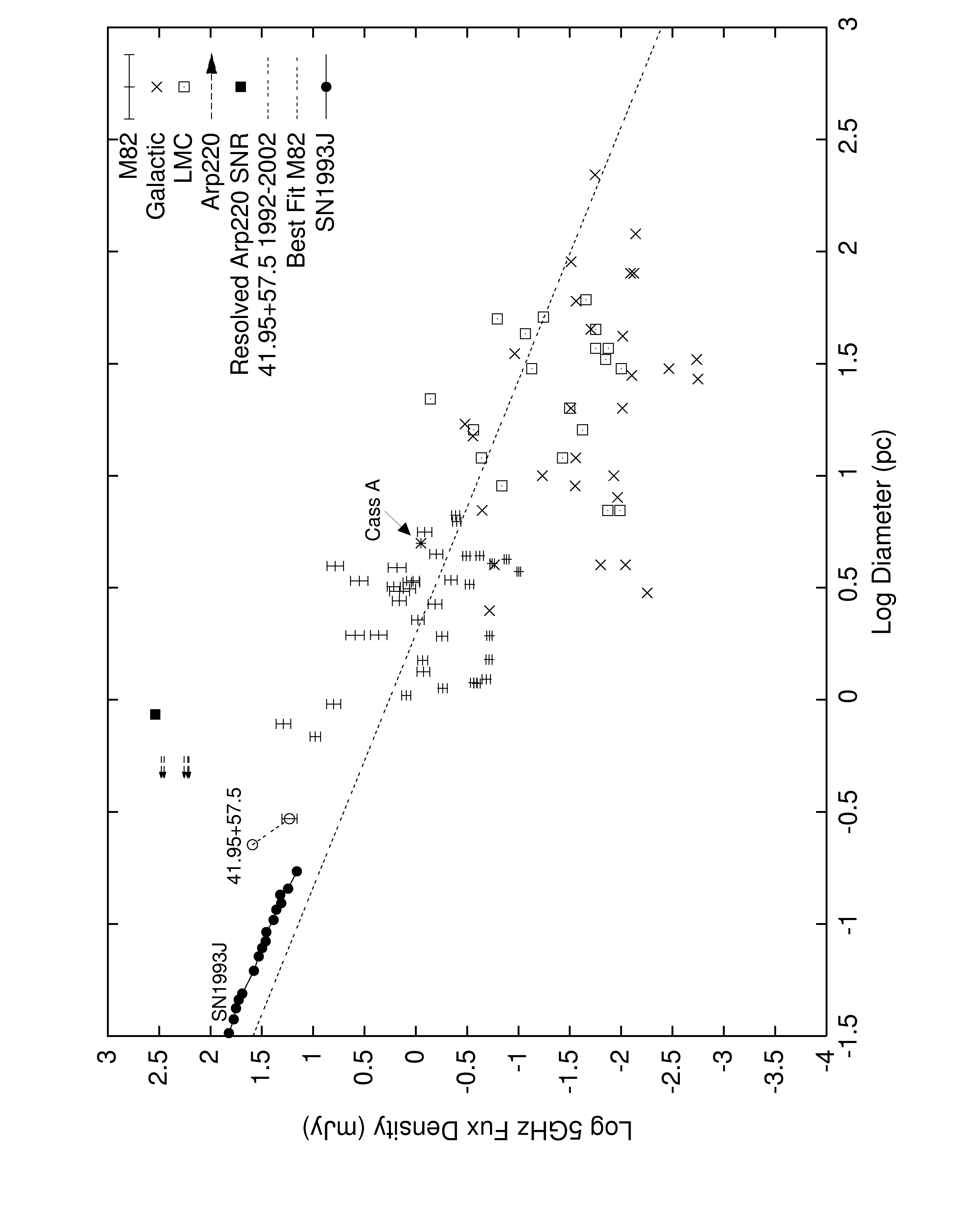}
\includegraphics[width=8cm,angle=270]{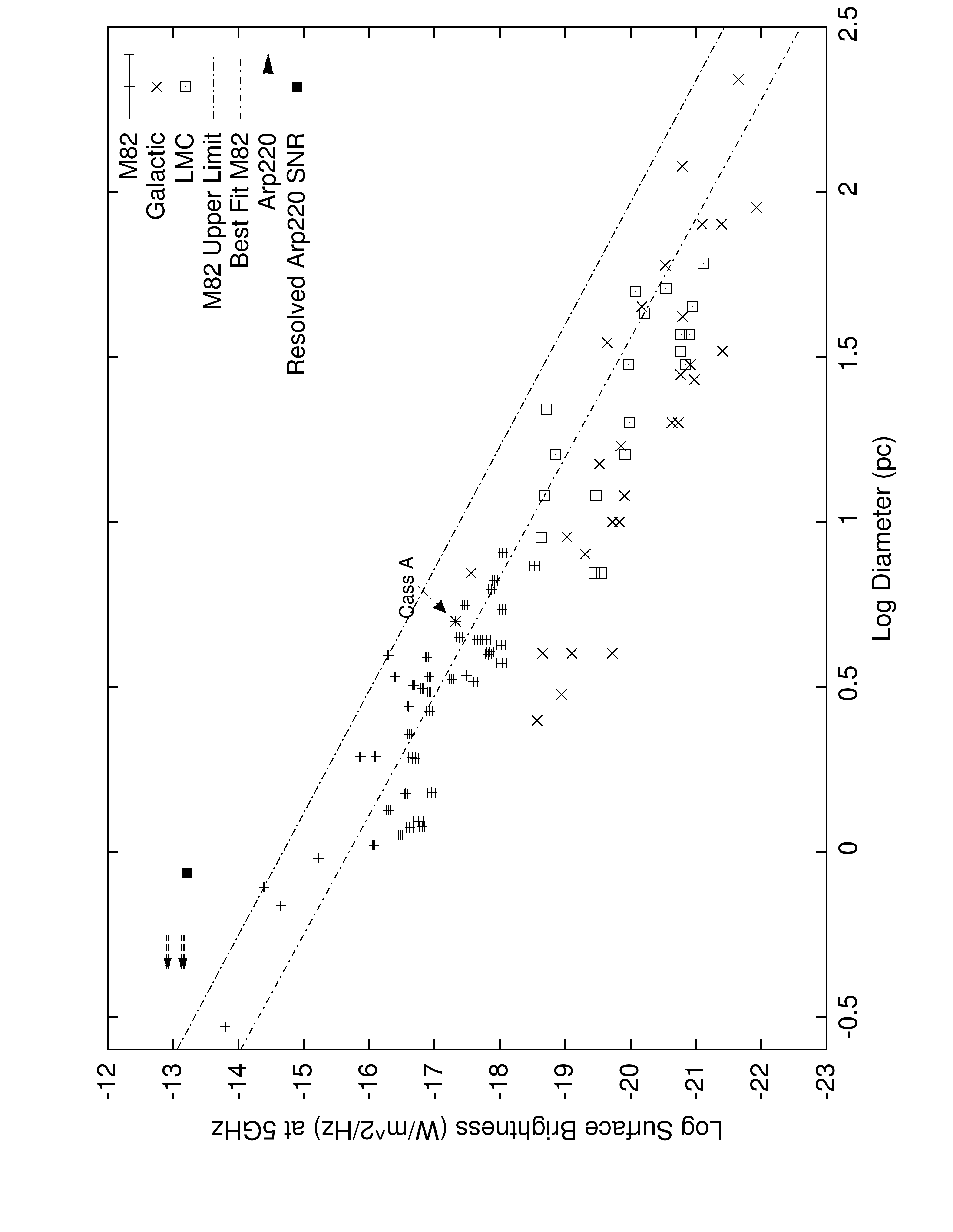}
\hfill
\end{minipage}
\caption{The SNR 5\,GHz Flux density and surface brightness as a function of diameter. Galactic \protect{\citep{green04}} and LMC SNR \citep{mills84} have been included for comparison. The best-fitting line and upper limits for the M82 SNRs are included. The arrows (upper limits) and filled box (resolved) show Arp\,220 sources from \protect\cite{parra07}. The star shows the position of the Galactic remnant Cassiopeia A.The filled circles show 5\,GHz observations of SN1993J taken from \protect\cite{bartel02} from Nov. 1993 to Jun. 2001.}
\label{surfb}
\end{center}
\end{figure}

Fig. \ref{surfb} shows a plot of the 5\,GHz flux density (S), as a function of their diameter (D), for the M82 SNRs as well as a number of Galactic \citep{green04} and Large Magellanic Cloud (LMC) \citep{mills84} SNRs.  Also included are sizes and limits for sources detected by VLBI observations, including SN1993J \citep{bartel02} and SNR from Arp\,220 \citep{parra07}. The Galactic and LMC source flux densities have been extrapolated to 5\,GHz using the spectral indices presented in the literature, and the diameters of all the sources have been scaled to the distance of M82. For the M82 SNRs, there appears to be a relationship of the form, $S \propto D^{-0.9}$, (shown by the dotted line in Fig. \ref{surfb}), whereas the general trend for all of the SNRs indicates a somewhat steeper relationship. However, this latter trend is subject to considerable error because of the uncertainties in the distance estimates for the Galactic SNRs. The overlap in the samples plotted show that, whilst the M82 SNRs are more luminous and compact, they appear to belong to the same class of objects as the SNRs found in our own galaxy. 

This 5-GHz Flux Density-Diameter relationship is most strongly restricted at the lower limit which is heavily affected by the sensitivity of the observations. The increased sensitivity of the 2002 deep observations has reduced this effect and enabled a more accurate determination of this lower limit. 

It should be noted that up to 40 per cent of the flux of the more extended sources could have been resolved out by MERLIN because of its $\sim 12$\,km shortest baseline length. On the other hand, this spatial frequency filtering means that very extended background emission, which makes it difficult to measure flux densities of individual sources with the VLA, will have been mostly resolved out by MERLIN. Although the comparison with Galactic and other extragalactic SNRs will be somewhat affected by resolution, the effects on the observed scatter is likely to be comparably small. The continued decay in the flux density of the SNR, 41.95+57.5, between 1992 and 2002 (see section \ref{notes}) means that its position in the plots shown in Fig. \ref{surfb} is changing.
 
The SNR included from observations of the ultraluminous infrared galaxy Arp\,220 are clearly much more luminous than those found in M82. The sample includes seven sources taken from \cite{parra07}, six of which are plotted with upper limits in diameter as they are unresolved, and one of which has a measured size of 0.86\,pc. These seven sources are labelled as L class sources in \cite{parra07} as they are at least 11 years old and have very small or no measured decay in flux density. These sources are therefore cited as likely SNR within Arp\,220. 

\onecolumn
%\twocolumn

The surface brightness ($\Sigma$) of the 38 SNRs have also been plotted in Fig. \ref{surfb} as a function of their diameter, D, assuming that the largest angular size approximates the diameter where no clear shell structure has been identified. This `$\Sigma$-D' relationship has often been used to study SNR populations \citep{mills84, green04}, although it can give a false impression of a correlation (effectively plotting 1/$D^2$ versus D) and should therefore, be treated with caution. The uncertainties in distance measurements for Galactic SNRs also make its use in such an analysis inappropriate. However, as the relative distances to the M82 SNRs are known to within $\sim$1 per cent, this relationship can be sensibly considered and is therefore included. The observed scatter in the M82 SNR population, (unattributable to errors in the distance estimates) is at least partially explained by differences in the ambient density associated with each SNR, a hypothesis supported by the variety of expansion velocities observed (see section \ref{msre}).
The same selection effects that affect the 5\,GHz flux density plot also limit the accuracy of this $\Sigma$-D relation, in particular the sensitivity of the observations.
A best fit line to the M82 data indicates $\Sigma\propto\,D^{-3.0}$, as illustrated in Fig. \ref{surfb}. It is also possible to fit an upper limit to the SNR samples included here, of $\Sigma\propto\,D^{-2.7}$, which is less steep than the proposed upper boundary of $\Sigma\propto\,D^{-3.5}$ by \cite{berkhuijsen87}. As has previously been reported \citep{muxlow94,green04}, the observed trends of the $\Sigma$-D relation for SNRs in M82 are consistent with those derived for Galactic SNRs, but that the SNRs in Arp\,220 are clearly much more luminous with respect to their equivalent sizes than those in M82 or the Milky Way.

\subsection{The cumulative distribution and average deceleration parameter}\label{ND}

\begin{figure}
\centering
\includegraphics[width=8cm,angle=270]{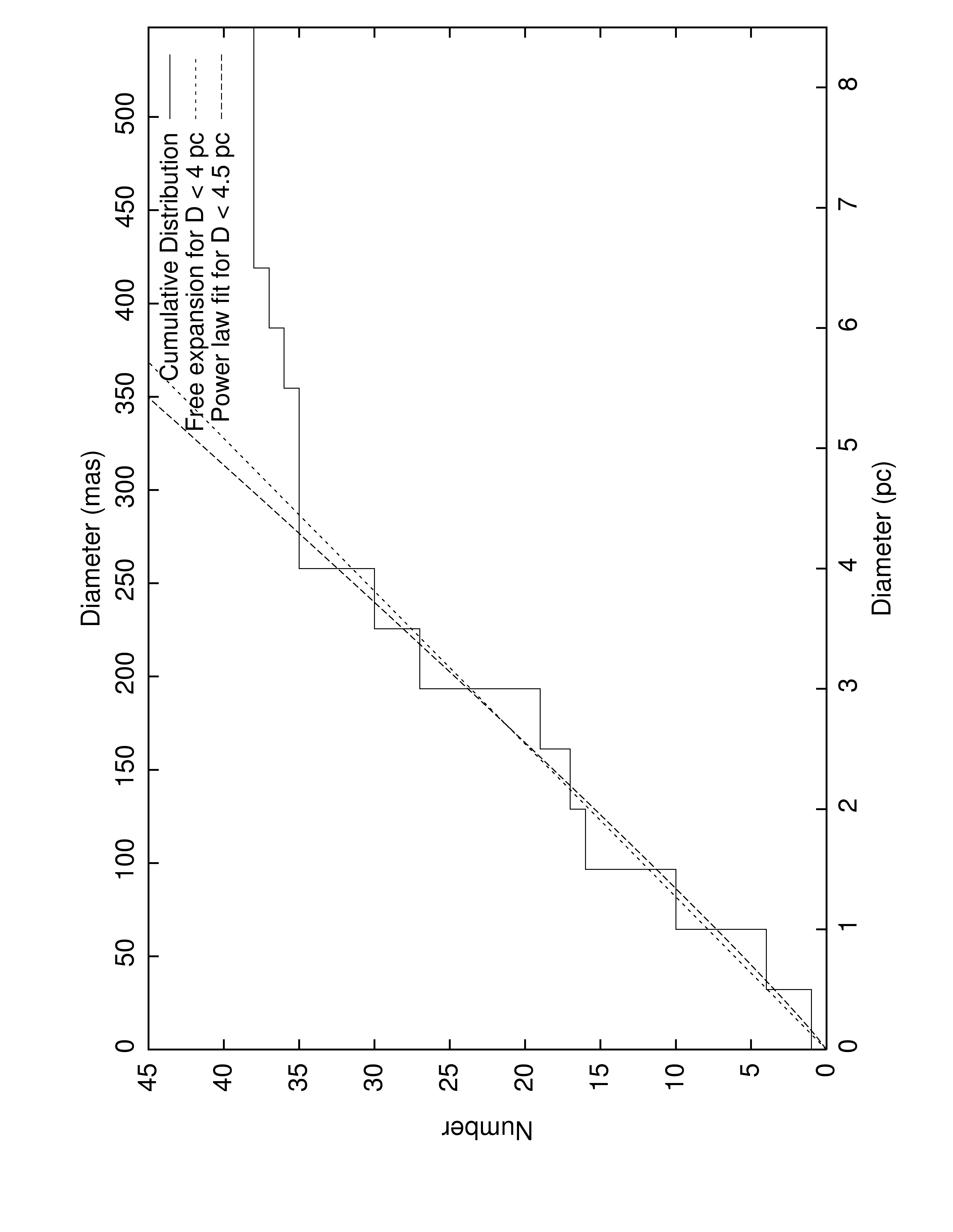}
\hfill
\caption{The cumulative distribution of the SNRs in M82. The large dashed line indicates the fit for SNRs following free-expansion for diameters $<4\,\rm{pc}$ and the small dashed line represents a power-law fit to calculate the deceleration parameter for diameters $<4.5\,\rm{pc}$.}
\label{cumdist}
\end{figure}

Fig. \ref{cumdist} shows the cumulative distribution of the sizes of the SNRs presented here. This can potentially give information on the supernova rate and SNR evolution and should follow the relation $N(<D) \propto \nu_{SN}D^{-1/m}$ \citep{huang94,muxlow94}, where $\nu_{SN}$ is the supernova rate and m is the average deceleration parameter. If it is assumed that the SNRs are in a uniform density medium and are undergoing free expansion, the cumulative number will show a linear increase. As can be seen, the cumulative distribution can be fitted by a linear increase in diameter for D$\leq$\,4\,pc. 
If a SNR has entered the Sedov phase of its evolution, its size will increase as $D\propto\,T^{0.4}$, where T is the age of the SNR. This predicts the cumulative number to increase as $D^{2.5}$.
%For diameters greater than $\sim$4.0\,pc, there is a possible up-turn observed for D$\sim$4.5\,pc which could be indicative of a deceleration in the expansion rate of the SNRs. 
The best-fit power law to our sample of SNRs for D$<$4.5\,pc shows the cumulative number is increasing as $\sim D^{1.15}$, giving a lower limit to the average deceleration parameter of 0.93$\pm$0.06. This would suggest the SNRs are still experiencing free expansions and have not entered the Sedov phase. In the specific case of 43.31+59.2, \cite{beswick06} have calculated a lower limit of $\sim$0.68 for the deceleration parameter from VLBI monitoring. \\
\indent No evidence has been found to suggest that Galactic SNRs  up to $\simeq$400\,yrs old have begun to slow in their expansion \citep{green04}, and as $\sim$\,50\,\% of the known remnants in M82 are $<$330 years old, it is unlikely that deceleration will actually be observed. However, the high interstellar medium pressures thought to be present in the centre of a starburst could begin to affect the expansion of a SNR earlier than would be observed for remnants in our own Galaxy. A good example of this is the supernova SN1993J in the galaxy M81, which has been shown to have experienced varying degrees of deceleration over the last 14 years. \cite{bietenholz03} have discussed the evolution of the radio shell of SN1993J, derived from VLBI observations. These show the expanding shell to have begun a deceleration as little as 30 days after the explosion, but then to have increased to an observed maximum between 300 and 1600 days with a deceleration parameter of $\sim$0.74, by which point its expansion velocity had decreased to $\sim$8900\,\kms. The deceleration of this SNR has subsequently decreased to a deceleration parameter of $\sim$0.85.\\
\indent The diffuse background emission observed in M82 has a flux density $\sim$0.5\,Jy and is likely to be from evolved supernova remnants. The observed turnover in the N($<$D)\,-\,D plot for diameters above $\sim$\,5\,pc is a consequence of the fact that the sample is incomplete for sources with low surface brightness, because the surface brightnesses of such large evolved SNRs match that of the observed extended background emission, making them difficult to detect. The N($<$D)\,-\,D plot presented here is the most complete at 5\,GHz to date. As the instrument sensitivity improves, it should become possible to detect the more diffuse and weaker SNRs within M82.

\section{The interstellar medium}\label{ISM}

\begin{figure}
\begin{center}
\includegraphics[width=5cm,angle=270]{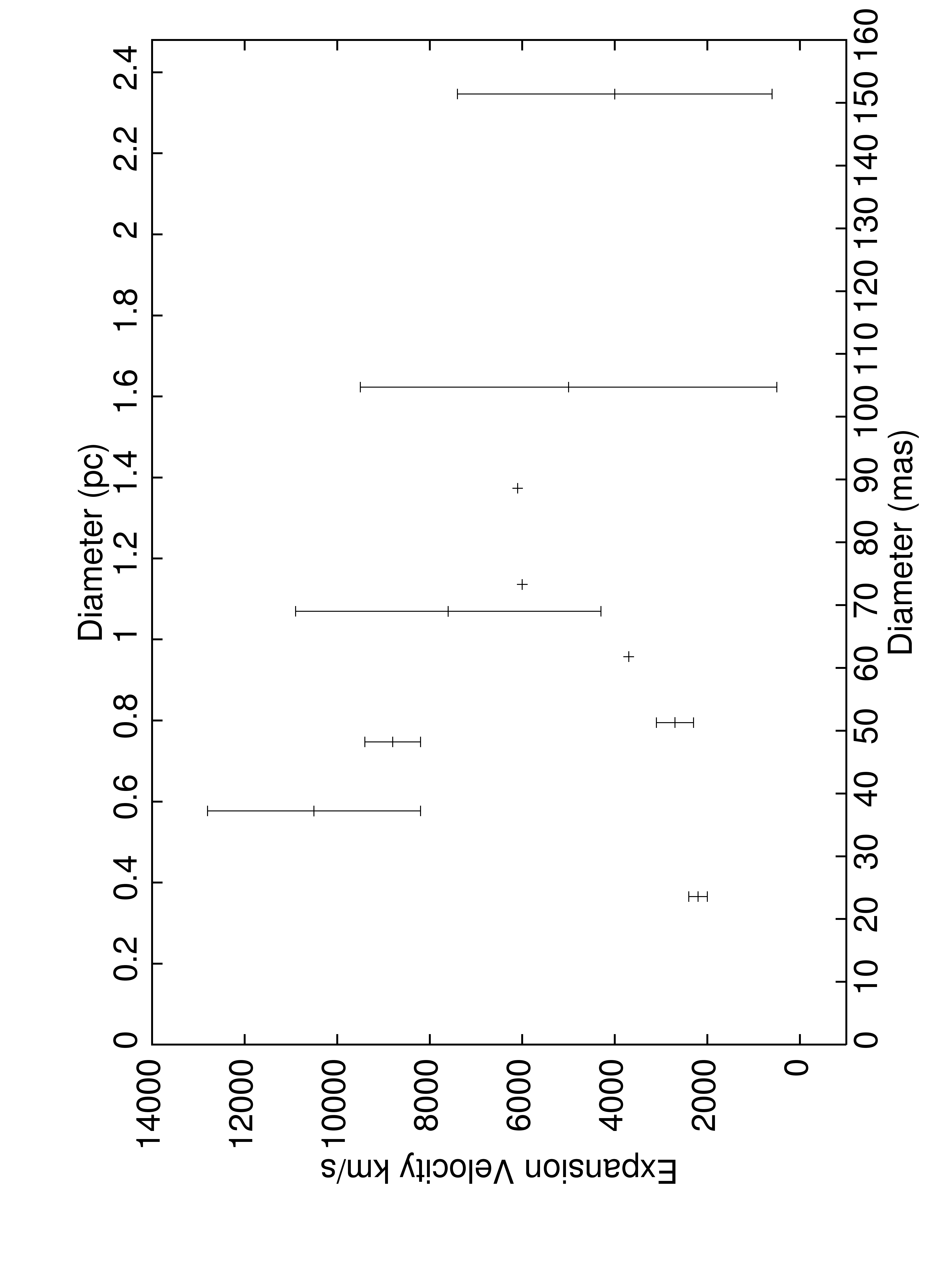}
\caption{The measured expansion velocities of ten SNR plotted against their diameter.}
\label{expvel}
\end{center}
\end{figure}

The range of expansion velocities observed (see section \ref{msre}) could initially be interperated as a consequence of the range in age of these sources. However, as can be seen in Table \ref{tab2}, the larger (and presumably older) remnants do not necessarily have slower velocities. Fig. \ref{expvel} shows a plot of expansion velocity versus source diameter. There appears to be no correlation, which would suggest that the SNRs are situated in very different, local gas density environments and are experiencing varying degrees of deceleration.

%H{\sc i} absorption studies of M82 \citep{wills98} have shown absorption against some of the remnants for which velocities have been presented here. Together with spectral index studies \citep[e.g.][]{mcdonald02} which show low frequency turnovers for some sources, it can be inferred that some of the remnants lie within or behind the gas fuelling the starburst. Conversely, remnants such as 43.31+57.3, which show no low frequency turnover, may be considered to lie on the near side of the starburst region. 
%The distribution of sources within the M82 ISM may partially explain the variety of expansion velocities seen. Thus, one might expect that remnants showing evidence of absorption should have lower expansion velocities because they are embedded within areas of higher local gas densities. However, some sources within the starburst region exhibit strong absorption and high expansion velocities, (such as 39.10+57.3). Such cases imply that the sources are situated on the far side of the starbursting region where the pressure of the surrounding region could be quite low, but that there is a large column density of foreground gas along the line of sight.\\\\

\indent \cite{chevalier01} assume a pressure in the starburst region of M82 of $10^7\,\rm{cm}^{-3}K$ and argue that the majority of the remnants observed are evolving into the interclump medium of molecular clouds with densities $\sim 10^{3}\,\rm{cm^{-3}}$. They use this to predict a canonical expansion velocity for the SNRs of $\sim$500\,\kms, which assuming that the sizes observed are typically a few parsecs, would imply ages of thousands of years for the SNRs in M82. These relatively low velocities are a factor of four smaller than the lowest observed velocities and a factor of $\sim$\,20 times smaller than the highest of the velocities just measured, as well as those measured from VLBI observations of the most compact sources \citep[e.g.][]{pedlar99,mcdonald01,beswick06}. Whilst there will be part of the M82 starburst at these pressures, it is evident from the size and spread of the velocities that the starburst region is not in a static pressure equilibrium and a range of pressures is present.\newline
\indent \cite{chevalier01} also suggest that the SNRs in M82 have entered the radiative phase of their evolution; as a remnant in a density of $\sim 10^{3}\,\rm{cm^{-3}}$ for explosion energies of $10^{51}$ ergs become radiative at a radius $\sim$ 1\,pc. The lower-limit to the average deceleration parameter calculated from the N($<$D)-D relation in section \ref{ND} of $\sim$0.93, would suggest that the observed population is still in the free-expansion phase of its evolution. However,  \cite{berkhuijsen87} suggest that a number of remnants evolving into an ISM with large density variations and experiencing varying degrees of deceleration could give rise to a N($<$D)-D relationship that looks similar to free-expansion, thus providing an explanation for the variety of expansion velocities observed. \\
\indent Assuming that the sources are undergoing free expansion and have not yet entered the Sedov phase of their evolution, an estimate can be made of the densities of gas into which the SNRs are expanding. 
This will cause a peak in the radio emission at a diameter given by $D\simeq 8.2(M_{1}/n_{0})^{1/3}$\,pc where $M_{1}$ is the ejected mass (in units of 10\solmas) and $n_{0}$ is the hydrogen density ($\rm{cm^{-3}}$) \citep{chevalier82,huang94}. The smallest SNR observed in M82 (with the exception of 41.95+57.4) has an average diameter of 0.65\,pc. Assuming $M_{1}=0.5$ (i.e. an ejected mass of 5\,\solmas), provides a density of $\sim 10^3\,\rm{cm^{-3}}$. This would clearly suggest that the SNRs in M82 are expanding into regions of the ISM with lower densities than the $10^{3}\,\rm{cm^{-3}}$ of the ambient medium, (as larger sources give rise to lower density estimates). Such low densities could be the result of stellar winds from the massive progenitor stars having driven away most of the surrounding material, and thus forming low density bubbles, prior to the supernova explosions taking place.

\cite{kronberg00} have studied the variability of 24 of the remnants in M82 and from statistical arguments, have proposed that only a quarter of their sample exhibits the expected decays in flux density of $\simeq$ 1\% per year. The remainder of the sample shows little variability, indicating radio-emitting lifetimes of $\sim$ 1000\,yrs. This would agree in part with the studies by \cite{chevalier01} suggesting that the expansion of the supernova remnants is much smaller ($\sim$500\,\kms) than those measured from the current observations. However, the velocities presented here not only confirm those made using independent VLBI investigations, but also provide robust measurements for an additional four SNRs; in all cases the observations show expansion velocities considerably in excess of the value of $\sim$500\,\kms suggested by \cite{chevalier01}.

\subsection{Supernova and star formation rates}\label{SFR}

It is possible to use the results presented here to directly estimate the supernova rate in M82. 
As indicated by the $\Sigma$-D relation discussed in section \ref{SNR}, Galactic and M82 SNR appear to be of the same class of object, and are therefore comparable. Cassiopeia A is an example of a Galactic SNR with a known age ($\sim330$ years). Assuming that the SNR in M82 that are more luminous than Cassiopeia A (22 of the total 37 detected) are younger, it is possible to estimate the supernova rate as: $\rm{\nu_{SN}\approx 22/330 \approx  0.07\,yr^{-1}}$.

\indent The supernova rate can also be calculated using the measured N($\rm{<D)-D}$ relation. Assuming that the sample is complete for D$<$4\,pc and that the SNRs are in free expansion, the supernova rate can be estimated to be $\nu_{SN}\approx0.08(V_{exp}/5000)$. Taking an average expansion velocity of $\sim$5000\,\kms gives an estimate in very good agreement with the previous calculation.  
This is clearly the case for the remnant, 43.31+592, which is taken to be representative of the whole population in the light of the calculated average deceleration parameter in section \ref{ND}.%, and for which images dating from 1972 \citep{kronberg75} are available. 

Assuming a Miller-Scalo Initial Mass Function (IMF), taking lower and upper mass limits of 0.1 and 100\,\solmas\, respectively, these supernova rates can be used to calculate a star formation rate (SFR) using \citep{condon92,cram98} $$SFR(M\geq5\,\rm{M_\odot})=25\,\nu_{SN}\rm{M_\odot}yr^{-1}.$$ Using the above values for $\nu_{SN}$ this implies a SFR of $(SFR\geq5$\solmas$)\sim 1.8-2\,$\solmas\,$\rm{yr^{-1}}$ which is in close agreement with the value of $\sim$2\solmas\, derived from FIR, H$\alpha$ and Ultraviolet observations \citep[e.g.][]{young96,bell01}.

%Unlike the variability arguments of \cite{kronberg00} and the models of \cite{chevalier01}, we measure expansion velocities that imply ages of hundreds of years for the SNR in M82, in accordance with the supernova and star formation rates presented here, which are consistent with those derived from other tracers of star formation such as FIR.

\section{Conclusions}\label{conc}

An 8 day deep integration of M82 using the six element MERLIN array at 5\,GHz is presented. From this
\begin{enumerate}
\item{The sizes of 55 of the discrete sources, which include 37 supernova remnants with diameters ranging from 0.3 to 6.7\,pc, have been measured. The mean diameter of these supernova remnants is 2.9\,pc. The 13 sources identified as {\HII} regions have a mean diameter of 3.1\,pc with the full sample ranging from 0.7 to 6.5\,pc.}
\item{The observed SNR diameters follow the expected trends in the flux density and surface brightness versus diameter relationships seen for Galactic SNRs, showing the M82 SNRs to be of the same class as those found in the Milky Way, though probably younger.}
\item{Expansion velocities for ten SNRs have been derived using a previous epoch of MERLIN 5\,GHz observations made in 1992. This gives velocities ranging from 2200 to 10500\,\kms, which are in stark contrast to the predicted velocities of \cite{chevalier01} of $\sim$500\,\kms.}
\item{The cumulative number (N($\rm<D)-D$) relation has been used to calculate the average deceleration parameter for the SNRs in M82 which is found to have a lower limit of 0.93, indicating the SNRs are experiencing free expansion. This would appear to disagree with the measured expansion velocities which imply that the SNRs are undergoing deceleration at differing rates.}
\item{A supernova rate of  $\rm{\nu_{SN}\sim 0.07-0.08\,yr^{-1}}$ has been calculated using two independent methods. This has been used to derive a star formation rate of $\sim$1.8-2\,\solmas$\rm{yr^{-1}}$ in agreement with the SFR measurements in the FIR, H\,$\alpha$ and UV, which all provide rates of $\sim$2\solmas$\rm{yr^{-1}}$.}
\end{enumerate}

The measurement of the expansion velocities of ten SNRs (which until now has been limited to VLBI observations of the two most compact sources) enables the study of a SNR population without the limitations of similar investigations of SNRs in our own Galaxy. A combination of this deep MERLIN integration with future more sensitive observations of M82 should provide an excellent opportunity to measure the expansion velocities of a large sample of the supernova remnants within the central starburst. %The range of measured velocities is thought to indicate a spread of gas densities surrounding the SNRs, and thus a range in the experienced deceleration. Continued observation of these SNRs should also provide direct measurement of the deceleration of individual sources, confirming these conclusions. 
More sensitive observations will also enable the larger, more diffuse remnants with surface brightnesses close to the background emission to be observed, providing a more complete sampling of the SNR population, as well as enable the study of the individual HII regions. %The current observations suggest that the ISM is not in a pressure equilibrium, and the calculated deceleration parameter for the whole population would suggest that the SNRs are within regions of lower gas density than those predicted in models like that of \cite{chevalier01}, a hypothesis that can be tested by future observations.

\section*{Acknowledgments}

MERLIN is a national facility operated by The University of Manchester on behalf of the Science and Technology Facilities Council (STFC). D. Fenech wishes to acknowledge funding from PPARC. ParselTongue was developed in the context of the ALBUS project, which has benefited from research funding from the European Community's sixth Framework Programme under RadioNet R113CT 2003 5058187.

\end{document}